\documentclass[12pt,twocolumn]{aastex63}

\usepackage{xcolor}
\usepackage{cancel}
\usepackage{physics}
\usepackage{multirow}

\submitjournal{ApJS 21 Dec 2022 \\ Accepted for publication 28 Apr 2023}
\shorttitle{AGILE Observations of Solar Flares}
\shortauthors{Ursi et al.}
\newcommand{\cmmnt}[1]{\ignorespaces}

\begin{document}

\title{The First AGILE Solar Flare Catalog}

\correspondingauthor{Alessandro Ursi}
\email{alessandro.ursi@inaf.it}

\author[0000-0002-7253-9721]{A. Ursi}
\affiliation{\scriptsize INAF/IAPS, via del Fosso del Cavaliere 100, I-00133 Roma (RM), Italy}
\affiliation{\scriptsize ASI, via del Politecnico snc, I-00133 Roma (RM), Italy}

\author[0000-0002-4535-5329]{N.~Parmiggiani}
\affiliation{\scriptsize INAF/OAS, via Gobetti 101, I-40129 Bologna (BO), Italy}

\author[0000-0002-5422-1963]{M.~Messerotti}
\affiliation{\scriptsize Osservatorio Astronomico di Trieste, via G. B. Tiepolo 11, I-34131 Trieste (TS), Italy}

\author[0000-0002-4590-0040]{A.~Pellizzoni}
\affiliation{\scriptsize INAF - Osservatorio Astronomico di Cagliari, via della Scienza 5, I-09047 Selargius (CA), Italy}

\author[0000-0001-6661-9779]{C.~Pittori}
\affiliation{\scriptsize SSDC/ASI, via del Politecnico snc, I-00133 Roma (RM), Italy}
\affiliation{\scriptsize INAF/OAR, via Frascati 33, I-00078 Monte Porzio Catone (RM), Italy}

\author[0000-0003-2501-2270]{F.~Longo}
\affiliation{\scriptsize Dipartimento di Fisica, Universit\`a di Trieste and INFN, via Valerio 2, I-34127 Trieste (TR), Italy}

\author[0000-0003-3455-5082]{F.~Verrecchia}
\affiliation{\scriptsize SSDC/ASI, via del Politecnico snc, I-00133 Roma (RM), Italy}
\affiliation{\scriptsize INAF/OAR, via Frascati 33, I-00078 Monte Porzio Catone (RM), Italy}

\author[0000-0002-6230-665X]{A.~Argan}
\affiliation{\scriptsize INAF/IAPS, via del Fosso del Cavaliere 100, I-00133 Roma (RM), Italy}

\author[0000-0001-6347-0649]{A.~Bulgarelli}
\affiliation{\scriptsize INAF/OAS, via Gobetti 101, I-40129 Bologna (BO), Italy}

\author[0000-0003-2893-1459]{M.~Tavani}
\affiliation{\scriptsize INAF/IAPS, via del Fosso del Cavaliere 100, I-00133 Roma (RM), Italy}
\affiliation{\scriptsize Universit\`a degli Studi di Roma Tor Vergata, via della Ricerca Scientifica 1, I-00133 Roma (RM), Italy}

\author{P.~Tempesta}
\affiliation{\scriptsize Telespazio SpA, Centro Spaziale del Fucino 23, Piana del Fucino, Via Cintarella, I-67050, Ortucchio (AQ), Italy}

\author{F.~D'Amico}
\affiliation{\scriptsize ASI, via del Politecnico snc, I-00133 Roma (RM), Italy}


\begin{abstract}

We report the Astrorivelatore Gamma ad Immagini LEggero (AGILE) observations of solar flares, detected by the on board anticoincidence system in the 80-200~keV energy range, from 2007 May 1st to 2022 August 31st. In more than 15~yr, AGILE detected 5003 X-ray, minute-lasting transients, compatible with a solar origin. A cross-correlation of these transients with the Geostationary Operational Environmental Satellites (GOES) official solar flare database allowed to associate an intensity class (i.e., B, C, M, or X) to 3572 of them, for which we investigated the main temporal and intensity parameters. The AGILE data clearly revealed the solar activity covering the last stages of the 23rd cycle, the whole 24th cycle, and the beginning of the current 25th cycle. In order to compare our results with other space missions operating in the high-energy range, we also analyzed the public lists of solar flares reported by RHESSI and Fermi Gamma-ray Burst Monitor. This catalog reports 1424 events not contained in the GOES official dataset, which, after statistical comparisons, are compatible with low-intensity, short-duration solar flares.

Besides providing a further dataset of solar flares detected in the hard X-ray range, this study allowed to point out two main features: a longer persistence of the decay phase in the high-energy regime, with respect to the soft X-rays, and a tendency of the flare maximum to be reached earlier in the soft X-rays with respect to the hard X-rays. Both these aspects support a two-phase acceleration mechanism of electrons in the solar atmosphere.

\end{abstract}



\section{Introduction}
\label{introduction}

Solar flares represent intense and localized emissions of electromagnetic radiation occurring in the solar atmosphere, strictly correlated with the 11-yr solar cycle. These events are thought to be produced when accelerated charged particles interact with the plasma medium, releasing high-energy radiation in the X-ray regime. Solar flares can occur together with other sun-related phenomena, such as coronal mass ejections (CMEs) and solar particle events. The investigation of flare parameters, such as \textit{duration}, \textit{rise} and \textit{decay} times, \textit{peak flux}, and \textit{fluence} is useful to evaluate the theoretical expectations from different production mechanism models. Moreover, the comparison of soft X-ray (SXR, $<10$~keV) and hard X-ray (HXR, $>10$~keV) counterparts of solar flares can help shedding light on the possible correlation between the emissions in the softer and in the higher-energy regimes. Up to now, it has been empirically observed that solar flare light curves in the SXR band exhibit temporal profiles similar to those of the time integrated curves in the HXR band, suggesting that a casual relationship exists between the two components, called the Neupert effect \citep{Neupert1968,Dennis&Zarro1993}.

It is well known that the X-ray emission in solar flares can be ascribed to bremsstrahlung radiation produced by electrons accelerated in the solar atmosphere. It has been suggested that this acceleration process could consist in a first step, occurring in coincidence with the flash phase and injecting medium-energy ($<100$~keV) particles, and a second step, operating in a shock front and accelerating them to higher energies. Such two-phase mechanism would justify the large range of energies of the emitted particles, and would provide the sufficient time to make low-energy particles reach higher energies, even with a relatively slow acceleration process \citep{Bai&Ramaty1976,Bai&Dennis1985}.

Solar flares in the X-ray energy range are usually detected by a number of satellites for solar monitoring \citep[e.g., Geostationary Operational Environmental Satellites, hereafter GOES, RHESSI,][]{Veronig2002,Sui2004}, as well as for high-energy astrophysics \citep[e.g., Konus-Wind, Fermi,][]{Palshin2014,Ajello2021}. These events are of particular relevance, as they have an important role in space physics and space weather. Sudden bursts of radiation and accompanying solar wind, or CMEs, may affect the Earth's magnetosphere environment, possibly inducing damages on spacecraft electronics and on ground-based communication systems.


\section{The AGILE Satellite}

The Astrorivelatore Gamma ad Immagini LEggero (AGILE) satellite is a space mission of the Italian Space Agency (ASI), launched in 2007 and devoted to high-energy astrophysics \citep{Tavani2008b}. Its payload consists of an imaging gamma-ray silicon tracker (30~MeV--50~GeV), a coded mask X-ray imager SuperAGILE (SA, 18--60~keV), a nonimaging Mini-CALorimeter (MCAL, 0.4--100~MeV), and an anticoincidence (AC, 50--200~keV) system, composed of five independent plastic scintillation panels (four lateral and one top). AGILE is a low-Earth orbit (LEO) satellite, flying at $\sim500$~km altitude, in a quasi-equatorial ($2.5^{\circ}$ inclination) orbit. The AGILE data are downloaded at the ASI Ground Station in Malindi (Kenya) and then processed at the AGILE ASI-SSDC data center \citep{Pittori2019}, delivering data and scientific alerts for transients within 20~minutes - 2~hr since the on board acquisition.

\subsection{The Anticoincidence System}

The AGILE AC is a system of Bicron BC-404 plastic scintillation detectors, acting as charged particle detectors \citep{Perotti2006}. It includes a top panel (AC-Top), sensitive in the 50-200~keV energy range, and four independent lateral sides (AC-Lat1, AC-Lat2, AC-Lat3, and AC-Lat4), sensitive in the 80-200~keV energy range. Each lateral panel is composed of three independent detectors, making the whole AC system consist of 13 independent detectors. The AC data  are continuously recorded in telemetry as scientific RateMeter (RM) data, with a 1.024~s time resolution, together with data acquired by other on board detectors, such as SA and MCAL. The AC RM data provide a continuous monitoring of the X-ray background modulation along the satellite orbit, and typically reveal a large number of high-energy transients occurring on different timescales, from few seconds to tens of minutes, such as gamma-ray bursts (GRBs), flaring activity from soft gamma-ray repeaters (SGRs), and solar flares.

In this work, we focus on the AGILE monitoring of the Sun. Solar activity is best detected in the data acquired by AC-Lat4, whose surface is parallel to the spacecraft solar panels, always pointing toward the Sun and providing the best sensitivity to X-ray solar flares. On the other hand, the SA (18-60~keV) detector has a pointing direction that is orthogonal to that of the AC-Lat4: this means that the Sun is constantly outside its field of view, preventing any significant detection of solar flares. Only in some cases of particularly intense flares, signals have been spotted in the SA RMs, although they cannot be used to perform reliable analyses. No solar flares have been detected in the MCAL RMs (0.4-100~MeV): this lack can be ascribed both to the geometry of the detector, whose area is minimum at $90^{\circ}$ off axis, and to the rarity of events in the 0.4-100~MeV energy range (from the public list of RHESSI flares, it turns out that events detected in the $>300$~keV channels constitute 0.03\% of the whole sample). In this study, we only focus on solar flares detected with the AC-Lat4, in order to deal with a homogeneous sample of events, detected by a unique detector, under almost constant angular configurations and sensitivity.

The AGILE team developed a real-time pipeline to routinely scan the AGILE AC-Lat4 data stream, acquired in the 80-200~keV energy range, searching for minute-lasting, high-energy transients. This pipeline belongs to the AGILE RTA system \citep{Bulgarelli2019Ex}. If a candidate solar flare is detected, the system delivers an automatic alert to the AGILE Team, and builds up a comprehensive database of hard X-ray transients which can be used to monitor the solar activity. The AGILE pipeline allowed to promptly react to solar activity, and to communicate some remarkable flaring activity \citep[e.g.,][]{Ursi2020solar,UrsiGCN2022}.


\subsection{Issues to the AGILE Solar Monitoring}
\label{issues}

It is important to point out two main issues which affect the AGILE solar monitoring and which put important constraints to our observations:

\begin{itemize}

\item The existence of time intervals during which the observation of the Sun is prevented (typical of LEO orbits). This issue affects the capacity of a satellite to detect a solar flare in its entirety, preventing precise characterization of the duration of the events. As AGILE is an LEO satellite, it experiences a systematic number of Earth occultations (EOs) of the Sun and passages into the South Atlantic Anomaly (SAA), where all detectors are usually put into an idle mode, due to the massive presence of high-energy charged particles. For what concerns EOs, an LEO satellite typically experiences an ``umbra'' region (i.e., the orbital region in which the Sun is not visible at all) consisting in about 37\% of its entire orbital path. As a consequence, out of a total orbital period of 100~minutes, AGILE travels about $\sim37$~min in the Earth's \textit{night-time} side, where no observation of the Sun is possible. On the other hand, this region reduces to only 5\% for a geostationary satellite, pointing out the best suitability of that orbit to solar monitoring. For what concerns the SAA, as the AGILE inclination is very small, the size of the crossed geographic region interested by the presence of high-energy charged particles is quite constant during the orbit, corresponding to an average time interval of $\sim10$~minutes out of 100~minutes. Throughout the orbits, as the Earth rotates, the EO region precedes and may fully, partially, or not at all overlap with the SAA region. This translates into exposures to the Sun going from 63~minutes (best case) to 53~minutes (worst case). We recall that these are average values which may vary depending on the satellite altitude and the time of the year. Solar flares lasting longer can be revealed, but they will be more likely affected by data gaps. Considering the different EO and SAA regions encountered throughout the orbits, we end up with a mean 55\% time of AGILE exposing the Sun, pointing out that about half of the events produced by the Sun may not be detected by a typical LEO satellite.

\item The saturation of the detector count rate for very high fluxes. This issue affects the evaluation of the solar flare intensity, preventing a reliable characterization of the position of the \textit{maximum} inside the light curve. As the AC RM count rate saturates at 48,960~counts~bin$^{-1}$, the transients releasing higher count rates cannot be distinguished. This affects the possibility of evaluating the \textit{rise} and \textit{decay} times, the \textit{peak counts}, and the \textit{integrated counts} of the events. This issue mainly regards the most intense solar flares (classes M and X), often preventing the precise flux characterization of these events.

\end{itemize}


\section{The GOES Satellites}
\label{GOES}

We made use of data acquired by the GOES of the National Oceanic and Atmospheric Administration, which typically detect solar flares in the SXR energy range. GOES data are useful to carry out a cross-correlation with the AGILE transients, in order to check and confirm these events as known, genuine solar flares.

The first satellite of the GOES constellation (i.e., GOES-1) was launched in 1975, whereas the last one (i.e., GOES-18) was launched in 2022. The disk-integrated soft X-ray emission measurements of the Sun carried out by GOES since 1976 allowed an almost continuous monitoring of the solar activity, fully covering solar cycles 21, 22, 23, and 24, as well as the current cycle 25. The flares detected by GOES are observed in the 0.1-0.8~nm wavelength band (i.e., 1.5-12.5~keV energy range). Flares are typically classified in five different classes by means of their peak intensity: each event is characterized by a letter (i.e., A, B, C, M, and X) and a number (from 1.0 to 9.9) which indicate the order of magnitude of the peak flux in a logarithmic scale (e.g., $An = n \times 10^{-8}$ W~m$^{-2}$, $Bn = n \times 10^{-7}$ W~m$^{-2}$, $Cn = n \times 10^{-6}$ W~m$^{-2}$, $Mn = n \times 10^{-5}$ W~m$^{-2}$, and $Xn = n \times 10^{-4}$ W~m$^{-2}$, with $n$ going from 0.1 to 9.9). In general, GOES satellites are capable of detecting flares with intensities above B1.0. In the GOES detection logic, the \textit{start} time of a solar flare is identified whenever a detected transient satisfies several conditions: (1) in the 1-minute time resolution data, the flux exceeds the B1.0 threshold (i.e., $10^{-7}$ W~m$^{-2}$) for at least four consecutive minutes; (2) within these four minutes, the flux is monotonically increasing; (3) the flux in the fourth minute is greater than 1.4 times the flux in the first minute. On the other hand, the \textit{end} time is reached when the flux returns to half the peak flux above the pre-flare level. Within the \textit{start} and \textit{end} times, the bin with the highest flux is classified as the \textit{peak flux} of the solar flare. For this study, we used the GOES official solar flare list available at \textit{ftp://ftp.swpc.noaa.gov/pub/warehouse/}.


\section{Methods}

We built up a comprehensive database of all the AGILE AC-Lat4 RM data, going from the beginning of 2007 May 1st to the end of 2022 August 31st. Between the end of 2012 and the beginning of 2013, AGILE was not operational for 4 months, resulting in an effective satellite lifetime to 180~months. For this whole time interval, we retrieved all the AGILE EOs, in order to remove all the noninvestigable time frames from our data stream: the sum of all these intervals, together with the sum of all the SAA intervals during which the satellite was not acquiring data, corresponds to about $47\%$ of the whole mission lifetime, resulting in a total solar exposure equal to $\sim8$~yr out of 15.

We carried out an extensive search for X-ray transients in the AC-Lat4 data stream, identifying all signals exceeding a threshold of $3\sigma$ above the background rate, for at least 60 consecutive 1.024~s bins ($\sim1$~minute). We required the detected transients to last at least $\sim1$~minute, in order to reject all the spikes produced by high-energy particles vetoed by the system, as well as to reject other shorter-duration high-energy transients, such as GRBs or SGR flares. For each detected transient, the first bin exceeding the threshold is marked as the \textit{start} and the final bin as the \textit{end} of the flare. We classify as \textit{maximum} the bin within the flare \textit{start} and \textit{end} times, in which the highest count rate is reached. The time at which the \textit{maximum} occurs is the \textit{maximum} time, whereas the count rate reached in that bin represents the \textit{peak counts} (proportional to the \textit{peak flux}). The time difference between the \textit{maximum} time and the \textit{start} time is denoted as \textit{rise} time, whereas that between \textit{end} time and \textit{maximum} time as \textit{decay} time. The sum of all counts released by the flare within the \textit{start} and \textit{end} times constitutes the \textit{integrated counts} (proportional to the \textit{fluence}).


\section{Results}

The search for transients in the AGILE AC-Lat4 database ended up with the identification of 5003 events. This sample contains all events that fulfilled the conditions imposed by the detection algorithm. We refer to this sample as \textit{level 0} sample.

Although these events were detected by the AC-Lat4 panel, which directly faces the Sun and exhibits minute timescales, we cannot confidently address all of them as genuine solar flares, since the AC system has no imaging capabilities. We therefore carried out a cross-search, using GOES data, in order to find out how many AGILE events have a GOES correspondence. The publicly available GOES flare data include several information, such as the \textit{start}, \textit{maximum}, and \textit{end} times of each detected event, together with information about the intensity class and the emitted \textit{fluence} in the 0.1-0.8~nm (1.5-12.5~keV). The cross-search associates a GOES flare to an AGILE transient whenever the AGILE \textit{maximum} falls within the GOES \textit{start} and \textit{end} times. We point out that, in case of count rate saturation, the position of the AGILE \textit{maximum} is not clearly distinguishable in the light curve and the association is considered fulfilled if at least one of the saturated bins falls inside the GOES window. This cross-search resulted in 3579 matches.

\begin{figure*}
\centering
\includegraphics[width=1.0\linewidth]{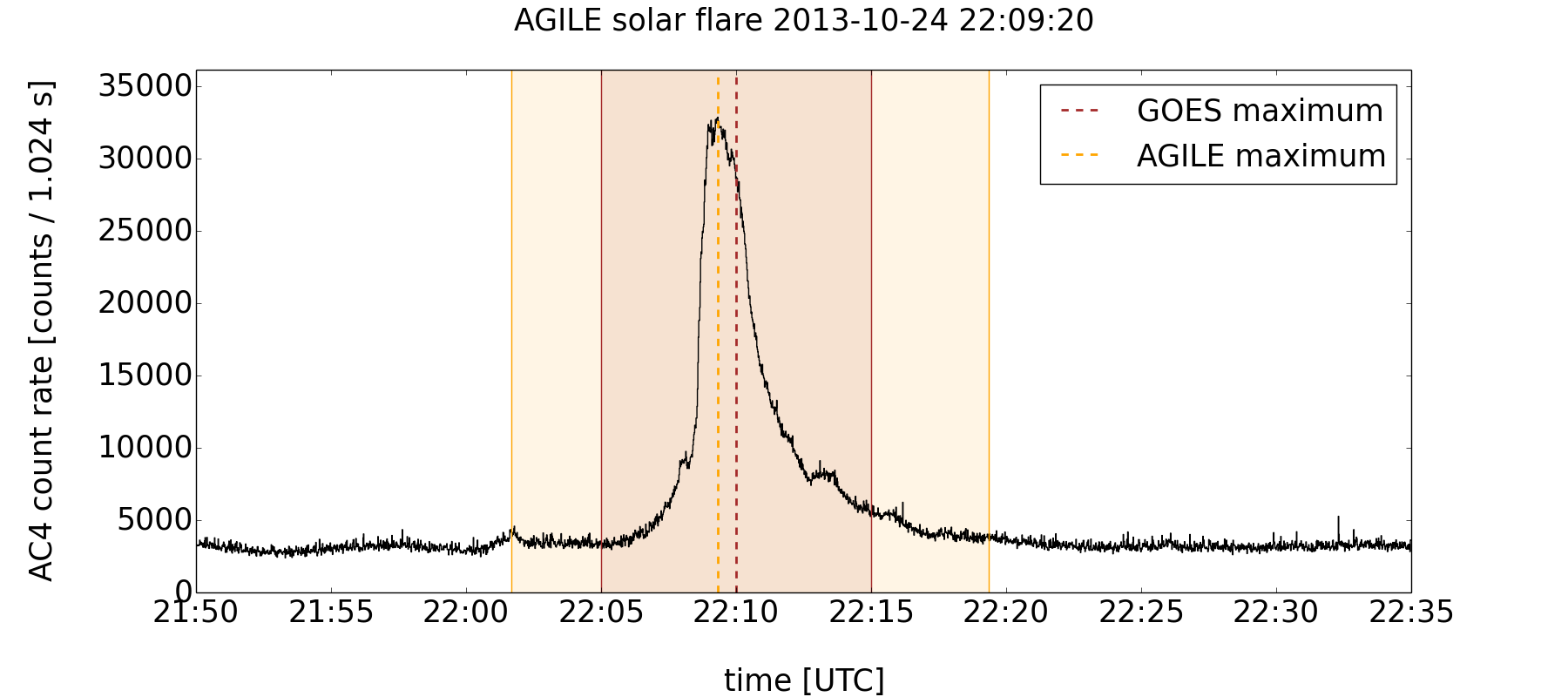}
\caption{Light curve of the AGILE AC-Lat4 for a solar flare detected on UT 2013 April 24 22:09:20. This transient, whose AGILE detection interval and \textit{maximum} are displayed in orange, is associated with a GOES flare of class C5.7, whose detection interval and \textit{maximum} are displayed in red. As the AGILE \textit{maximum} falls inside the GOES interval, the association is considered fulfilled and the event is classified as a genuine flare.}
\label{lc}
\end{figure*}

Fig.~\ref{lc} shows an example of a transient detected by the AGILE AC-Lat4 on UT 2013 April 24 22:09:20, which had been associated to a GOES flare of class C5.7. The association is fulfilled because the AGILE \textit{maximum} falls inside the GOES flare interval.


\subsection{Chance Matches}

A fraction of the obtained matches could arise from chance coincidences, given the large number of AGILE and GOES events. In order to evaluate the statistical robustness of the obtained GOES+AGILE associations, we carefully considered the GOES flare rate and the AGILE transient rate encountered in the time interval under analysis. The average duration of the GOES time intervals for the $N_{GOES}=20160$ solar flares reported in the GOES database is $\Delta t\sim13$~minutes. The total number of AGILE transients detected in a time interval of $T=180$~months is $N_{AGILE}=5003$. As a consequence, the number of expected chance matches is $N_{chance}=\frac{\Delta t_{GOES} \cdot N_{GOES}}{T} \cdot N_{AGILE} \sim170<<3579$. This means that only about 4\% of the matches could arise from chance coincidence, not constituting a relevant fraction of the overall sample.


\subsection{Superimposition of Solar Flares}
\label{superimposition}

We point out that more solar flares, produced in different regions of the Sun's disk, may occur at the same time, producing overlaps in the corresponding light curves. A detailed cross comparison of the time windows of the AGILE transients and of the GOES solar flares allowed to identify 7 cases in which an AGILE event was taking place during $>1$ GOES flares, produced in different positions of the solar disk. As these events cannot be uniquely associated to a specific GOES event, they are rejected from the AGILE+GOES sample. In the end, we obtain $3579-7=3572$ transients whose solar nature and intensity class are confirmed.


\subsection{Data Levels}
\label{data_levels}

Out of the 5003 \textit{level 0} sample, including all transients fulfilling the detection algorithm, we ended up with 3572 \textit{level 1} events, that can be used to study the solar activity, the solar flare rate throughout the years, and the statistics of the different intensity classes (see Section~\ref{solar_flare_rate}). The 7 events pointed out in Section~\ref{superimposition}, for which a unique GOES solar flare association is not possible, are not part of the \textit{level 1} sample and are referred to as \textit{level d} data.

Out of the \textit{level 1} events, 2149 flares have not been affected by EOs or SAA, building up the \textit{level 2} sample, which can be used to investigate the \textit{duration} of the detected solar flares, as the \textit{start} and \textit{end} times are clearly distinguishable (see Section~\ref{solar_flare_temporal_parameters}). Finally, out of them, 1983 solar flares did not saturate the AC-Lat4 count rate, constituting the \textit{level 3} sample, which provides a comprehensive sample to investigate also the \textit{rise} and \textit{decay} times (since the position of the \textit{maximum} is clear), the temporal offsets between AGILE and GOES, the \textit{peak counts}, and the \textit{integrated counts} (see Section~\ref{solar_flare_peak_flux_and_fluence}). We point out that \textit{level 3} sample is the most complete database that can be used for studies on the AGILE solar flares, as it consists of genuine events confirmed by matches with the GOES data, fully acquired on board, and whose count rate can be evaluated for the entire event duration.

\begin{figure*}
\centering
\includegraphics[width=0.8\linewidth]{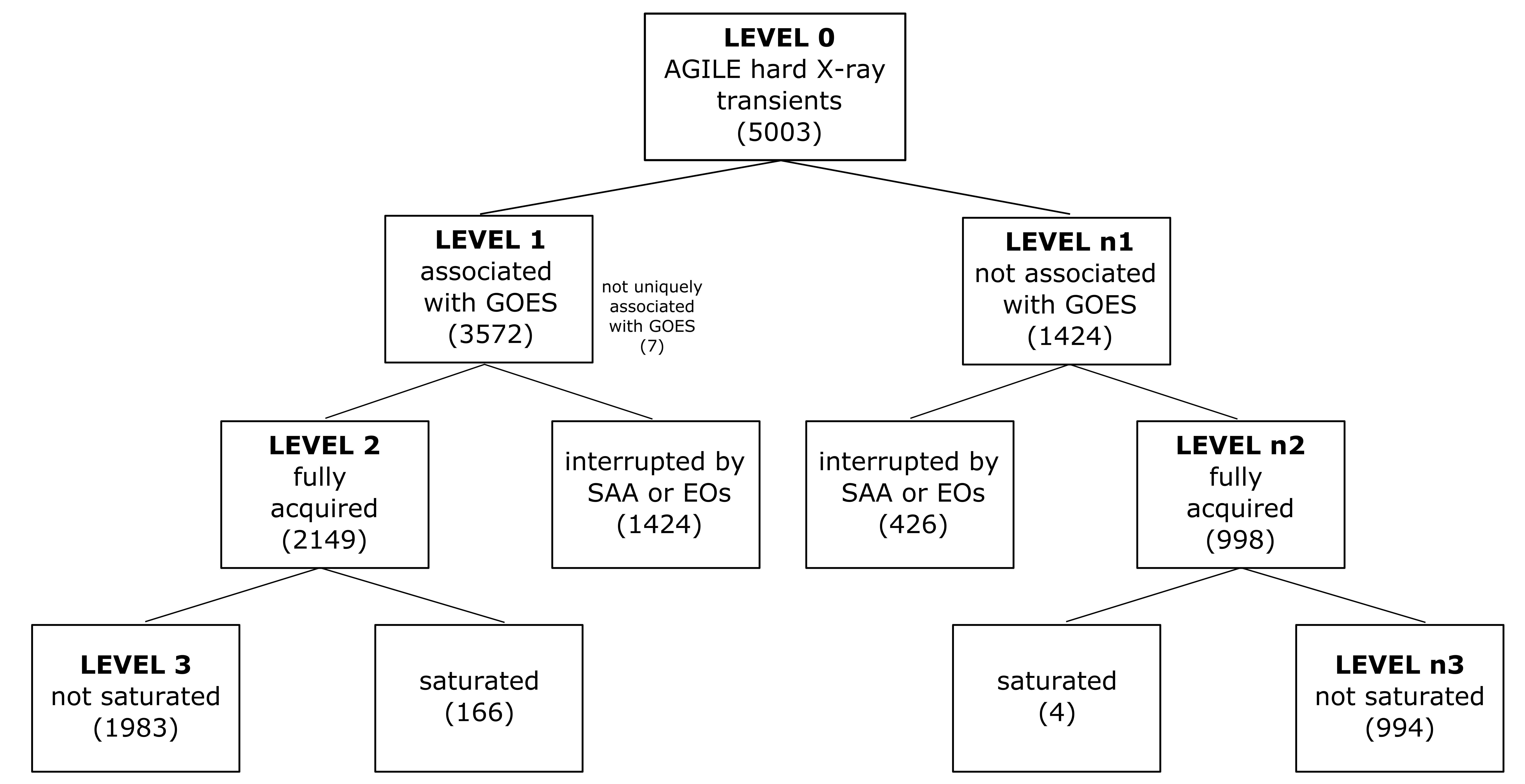}
\caption{Schematic representation of the results obtained in this work. Out of the 5003 transients (\textit{level 0}) obtained by the search algorithm for minute-lasting X-ray transients in the AGILE AC-Lat4 data, 3572 exhibited a unique associated GOES solar flare (\textit{level 1}). Out of these events, 2149 flares have been fully acquired on board (\textit{level 2}), and out of them, 1983 did not saturate the AC count rate (\textit{level 3}). On the right side, the events that did not match any GOES flare are classified into similar levels, on the basis of the same requirements (\textit{level n1}, \textit{level n2}, and \textit{level n3}).}
\label{events}
\end{figure*}

Similarly, the events that did not match any GOES flare constitute the \textit{level n1}, \textit{level n2}, and \textit{level n3} samples, respectively, classified in the same way as the matched flares (see Section~\ref{other_transients_detected_by_agile}). A scheme of the data levels is reported in Fig.~\ref{events}.

\subsection{Solar Flare Rate}
\label{solar_flare_rate}

\begin{figure*}
\centering
\includegraphics[width=0.7\linewidth]{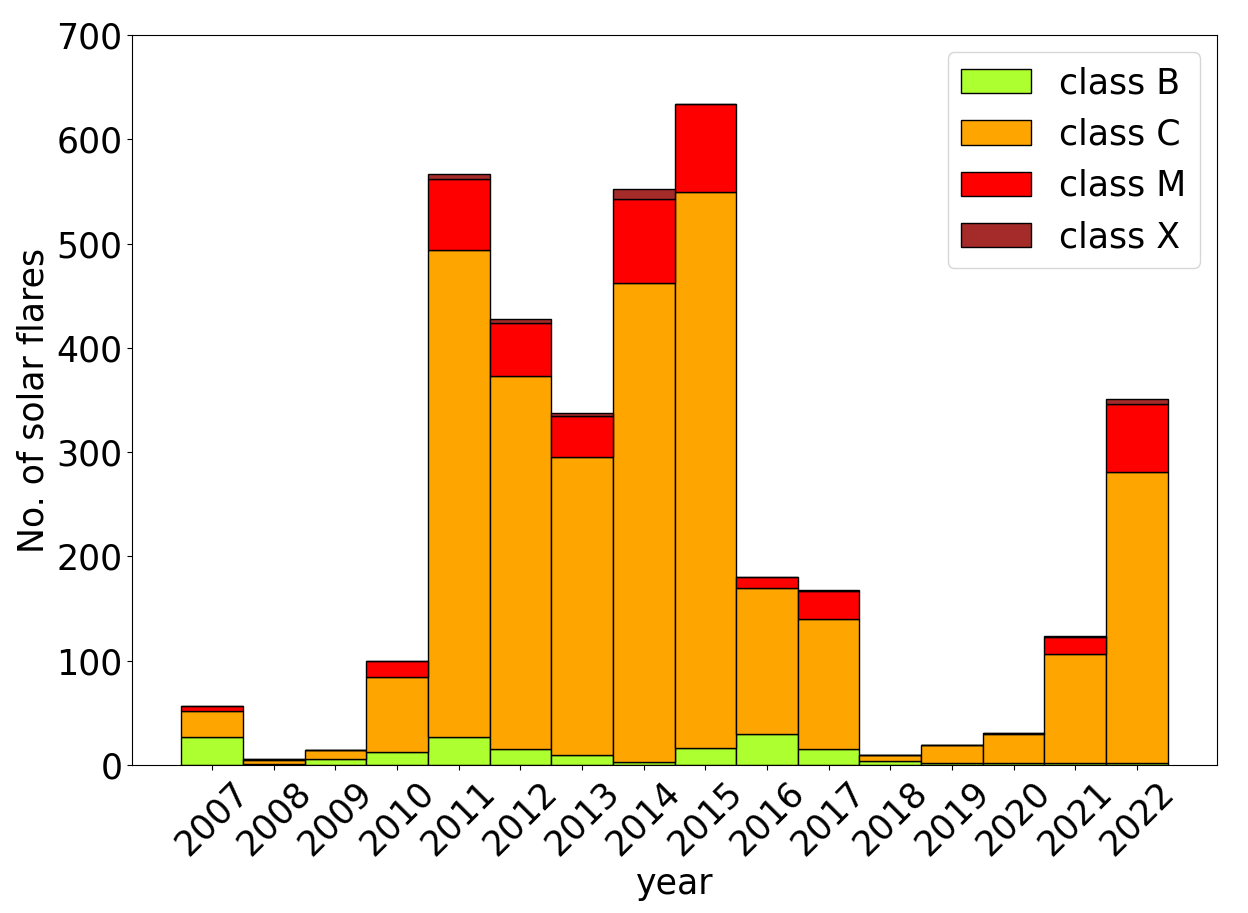}
\caption{Occurrence rate of the 3572 \textit{level 1} AGILE solar flares, detected by the AGILE AC-Lat4 between 2007 May and 2022 August. The end of the 23rd solar cycle (2007-2008), the whole 24th solar cycle (2008-2019), and the beginning of the current 25th cycle (2019-2022) are clearly revealed. Different colors denote different flare classes, as obtained from GOES.}
\label{histogram}
\end{figure*}

\begin{table}
\centering
\caption{AGILE solar flares}
\label{classes}
\begin{center}
\begin{tabular}{|c|c|c|c|}
\hline
\textbf{class}   & \textbf{Level 1} & \textbf{Level 2} & \textbf{Level 3}  \\
\textbf{(from}   & \textbf{(with GOES}    & \textbf{(Fully}      & \textbf{(Not}   \\
\textbf{GOES)}   & \textbf{Association)}  & \textbf{Acquired)}     & \textbf{Saturated)}   \\ \hline \hline
B   & 174                 & 155                    & 155                  \\
C   & 2904                & 1848                   & 1800                 \\
M   & 466                 & 141                    & 28                   \\
X   & 28                  & 5                      & 0                    \\ \hline
TOT & 3572                & 2149                   & 1983                 \\ \hline
\end{tabular}
\end{center}
Number of AGILE transients confirmed as solar flares by cross-correlations with GOES data (\textit{level 1}), fully acquired on board (\textit{level 2}), and not saturating the AC-Lat4 count rate (\textit{level 3}). Values refer to the different associated intensity classes and to the total sample.
\end{table}

Using the \textit{level 1} sample, we evaluated the AC-Lat4 solar flare detection rate encountered from 2007 May to 2022 August, reported in Fig.~\ref{histogram}. It can be clearly noticed that the decaying phase of the 23rd cycle, occurring until 2008 January, the whole 24th solar cycle, occurring from 2008 January to 2019 December, as well as the beginning of the currently ongoing 25th cycle, started in 2019 December. The drop of events during the 24th cycle is ascribed to the 4 months between the end of 2012 and the beginning of 2013 in which the satellite was not operational.

We evaluate the fraction of AGILE flares belonging to each class, whose values are reported in Tab.~\ref{classes}. The classes referred to are classes reported by GOES. For what concerns the \textit{level 1} sample, the majority of the events belongs to class C, which represents 80.3\% of the total sample. On the other hand, only 0.8\% belongs to class X. The lack of B class flares, evident in the ongoing 25th solar cycle, may be ascribed to the higher background rate experienced by the AC-Lat4 in the last years, due to the orbital decay of the AGILE satellite, which may affect the detection of some lower-intensity transient. 


\subsection{Solar Flare Duration}
\label{solar_flare_temporal_parameters}

In order to study the \textit{duration} of the solar flares detected by AGILE, it is fundamental that the events are completely acquired on board: this ensures to clearly distinguish the \textit{start} and \textit{end} times of each event and to cover their whole \textit{duration}. We therefore selected the 2149 \textit{level 2} flares. This sample still contains saturation effects, which however do not affect the \textit{duration} evaluation.

\begin{figure*}
\centering
\includegraphics[width=1.0\linewidth]{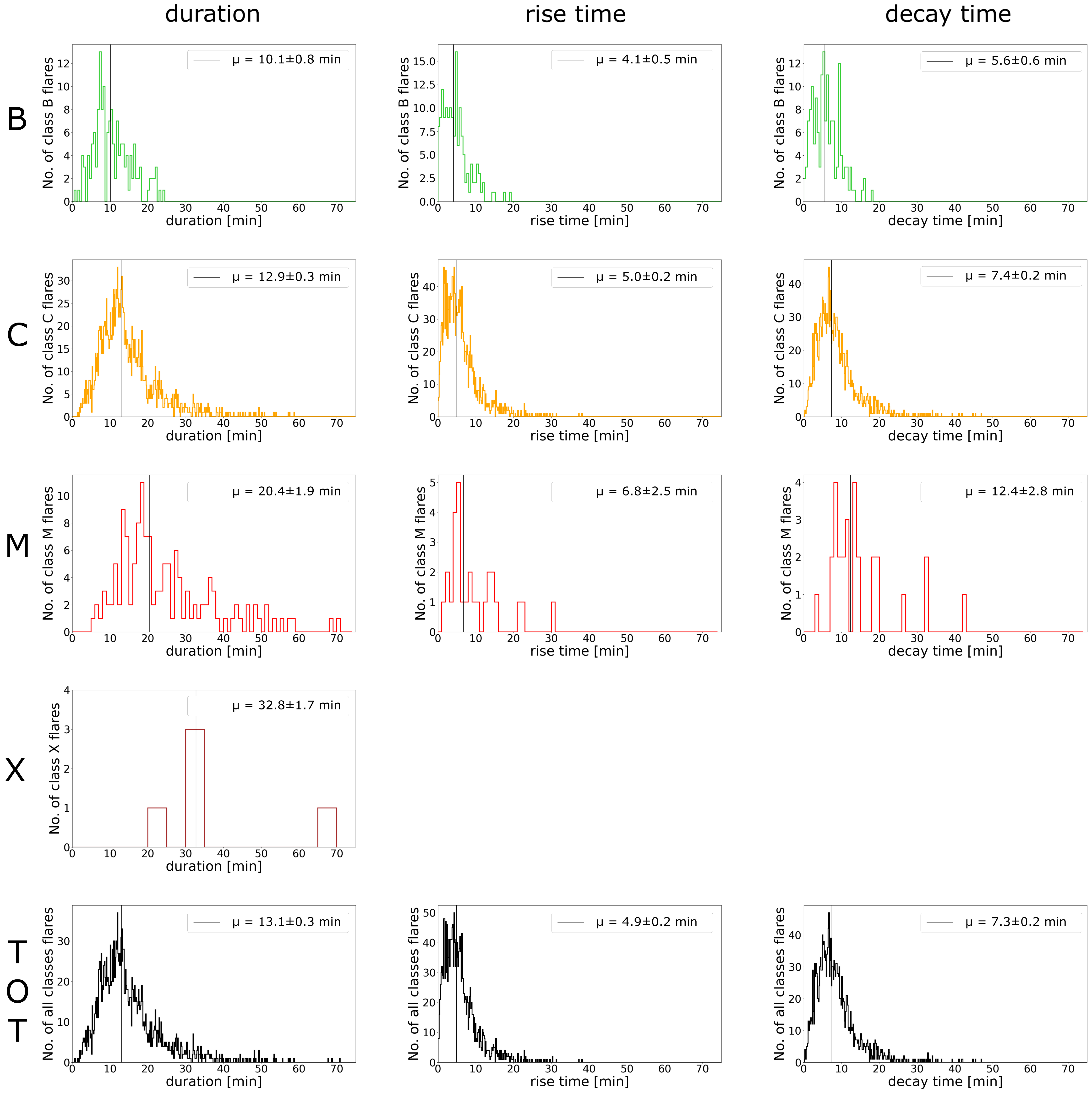}
\caption{Histograms of the \textit{duration} (first column), \textit{rise} time (second column), and \textit{decay} time (third column) of the solar flares detected by AGILE, divided into class B (first row), class C (second row), class M (third row), and class X (fourth row), together with the same distributions evaluated on the whole sample (fifth row). \textit{Duration} is estimated on the 2149 flares of the \textit{level 2} sample, fully acquired by AGILE, and is available for B, C, M, and X classes. On the other hand, \textit{rise} and \textit{decay} are evaluated on the 1983 flares of the \textit{level 3} sample, fully acquired and not suffering from count rate saturation, and are available for B, C, and M classes. For each parameter, the median value $\mu$ is reported as a vertical black line and its value is shown in the legend, expressed in minutes. B class and C class distributions are reported with a 30~s and 10~s bin-widths, respectively; on the other hand, M- and X class distributions are shown with 1 and 5~minutes, respectively, due to the poor statistics. The whole sample is evaluated with a 10~s bin-width.}
\label{distribution_duration}
\end{figure*}

\begin{table}
\centering
\caption{\textit{Duration}, \textit{Rise} Time, and \textit{Decay} time of the AGILE Flares}
\label{table_durationrisedecay}
\begin{center}
\begin{tabular}{|c|c|c|c|}
\hline
~ & \textbf{Level 2} & \multicolumn{2}{|c|}{\textbf{Level 3}} \\ \hline
\textbf{Class} & \textbf{Duration} & \textbf{Rise Time} & \textbf{Decay Time} \\
~ & \textbf{[minutes]} & \textbf{[minutes]} & \textbf{[minutes]} \\ \hline \hline
B & $10.1\pm0.8$ & $4.1\pm0.5$ & $5.6\pm0.6$ \\
C & $13.0\pm0.3$ & $5.0\pm0.2$ & $7.4\pm0.2$ \\
M & $20.4\pm1.9$ & $6.8\pm2.5$ & $12.4\pm2.8$ \\
X & $32.8\pm1.7$ & --- & --- \\ \hline
TOT & $13.1\pm0.3$ & $4.9\pm0.2$ & $7.6\pm0.2$ \\ \hline
\end{tabular}
\end{center}
Median values $\mu$~$\pm$~$c_{95}$ of the \textit{duration}, \textit{rise}, and \textit{decay} times obtained for the AGILE solar flares. \textit{Duration} is evaluated on the \textit{level 2} sample, whereas \textit{rise} and \textit{decay} times on the \textit{level 3} sample. Values refer to the intensity classes and to the total sample.
\end{table}

Tab.~\ref{classes} also reports the number of AGILE \textit{level 2} fully acquired solar flares, divided into each class, which represent the 60\% of the \textit{level 1} sample. This shows how, for an LEO satellite, the presence of interruptions due to EOs and SAA, although not preventing at all the detection of solar flares, severely affects the capacities of the satellite to fully acquire events on board, especially the longest ones. The \textit{level 2 / level 1} ratio becomes smaller with increasing intensity class. This could be ascribed to the general longer \textit{duration} of higher-class flares: these events may last several tens of minutes, being therefore more likely affected by interruptions due to EOs and SAA.

We evaluated the \textit{duration} of the \textit{level 2} solar flares and ended up with the distributions reported in the first column of Fig.~\ref{distribution_duration}. The first four rows correspond to the distributions evaluated on the events of the B, C, M, and X classes, respectively, whereas the last row is estimated on the total sample. In each panel, the solid line represents the median value $\mu$. These values are also reported in the first column of Tab.~\ref{table_durationrisedecay} as $\mu$~$\pm$~$c_{95}$, where the statistical significance equal to the 95\% confidence interval is $c_{95}=\frac{1.58(Q_3-Q_1)}{\sqrt{n}}$, with $Q_1$ and $Q_3$ indicating the first and third quartiles of the distribution, and $n$ is the total number of values. These distributions are compatible with those of typical GOES flares, reported by \cite{Veronig2002} for a sample of $\sim50,000$ events. Exploiting the subminute time resolution of the AGILE data, we report the B class distributions with a 30~s bin-width and the C class distributions with a 10~s bin-width; on the other hand, M- and X class distributions are shown with 1~min and 5~minutes, respectively, due to the poor statistics. The complete sample is distributed with a 10~s bin-width. As illustrated in Section~\ref{issues}, events lasting more than $\sim53$~minutes or $\sim63$~minutes (depending on the orbital configuration) are more difficult to be fully acquired on board, and represent only the 0.8\% and 0.1\% of the total, respectively. The shortest-duration AGILE solar flare lasted $0.8$~minutes (belonging to class B) and the longest one $70.8$~minutes (belonging to class M).

\subsection{Rise and Decay Times}
\label{rise_and_decay_times}

In order to study the \textit{rise} and \textit{decay} times, and the offset between AGILE and GOES detections, it is essential that the solar flare \textit{maximum} is uniquely identified in the corresponding light curve. This can be achieved only if the detector count rate does not saturate. We therefore select the 1983 \textit{level 3} events, to carry out the following analyses. The number of \textit{level 3} events, divided into each intensity class, is reported in the last column of Tab.~\ref{classes}. As expected from the definition of solar classes, the fraction of saturated events strongly increases with increasing flare class: while the totality of B class events did not saturate the AC, the totality of X class flares saturated it. The intermediate C- and M classes exhibit 97\% and 20\% of not saturated events, respectively: for what concerns C class flares, the saturated events have intensities above C5.0, whereas the M class events have intensities above M2.0. For what concerns the total sample, it is interesting to notice that the AC saturation affects about 8\% of the overall detected events.

We evaluated the \textit{rise} and \textit{decay} times, whose distributions are shown in the second and third columns of Fig.~\ref{distribution_duration}. The first four rows correspond to distributions estimated on the GOES B, C, and M classes, while the fifth row is evaluated on the total sample. The corresponding median $\mu\pm c_{95}$ values of \textit{rise} and \textit{decay} times are reported in the last two columns of Tab.~\ref{table_durationrisedecay}. It can be noticed that the median values of all temporal parameters tend to increase with increasing flare class. However, such increase is less evident in the \textit{rise} time, with respect to the \textit{decay} time, pointing out that the enhancement of the overall AGILE flares \textit{durations} is mostly ascribed to longer \textit{decay} times.


\subsection{AGILE-GOES Temporal Offsets}
\label{agile-goes_flare_offsets}

AGILE and GOES operate in different energy ranges and it is interesting to cross-correlate the temporal parameters as revealed by the two satellites. We evaluated the delay between the \textit{maximum} time of each flare, as detected by AGILE (in the HXR regime) and by GOES (in the SXR regime). 

\begin{figure}
\centering
\includegraphics[width=1.0\linewidth]{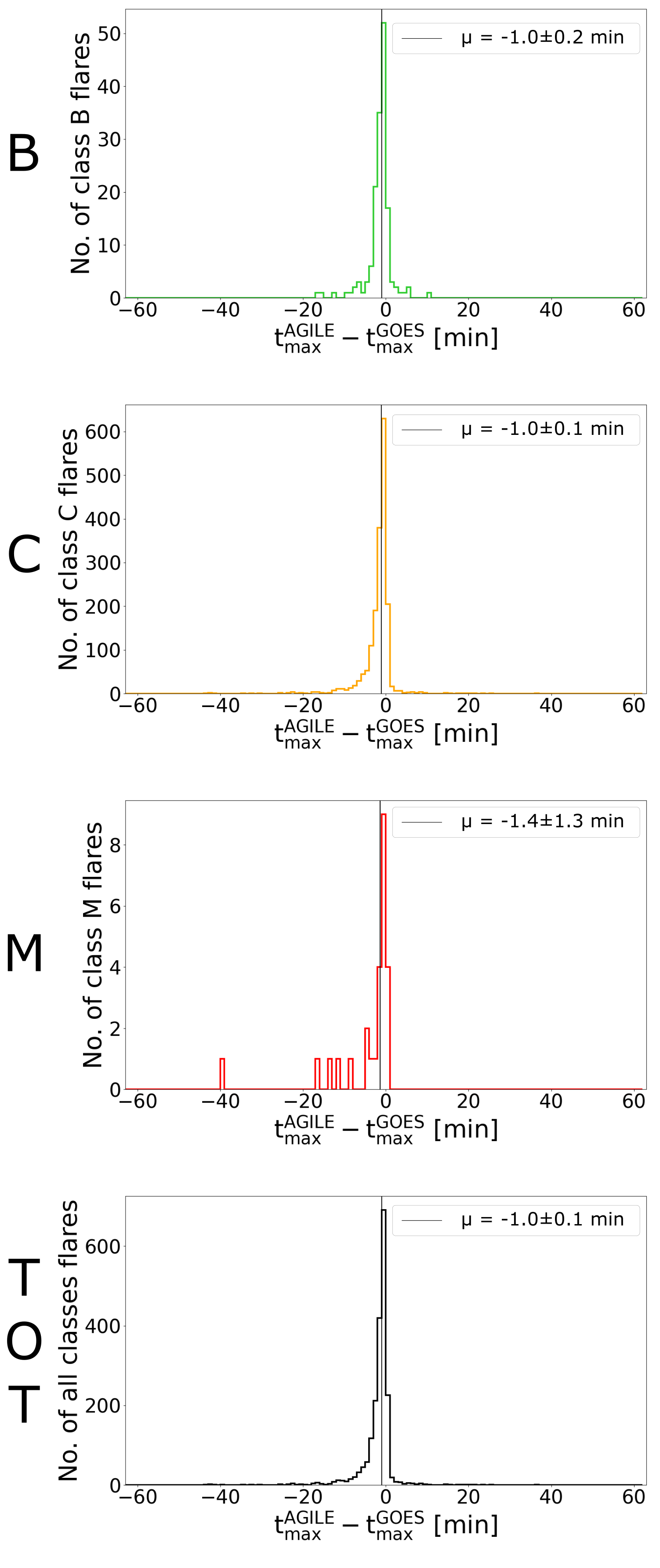}
\caption{Histograms of the offsets between the \textit{maximum} times, as seen in the AGILE and in the GOES data. The first three rows represent flares belonging to the B, C, and M classes, whereas the last row refers to the whole sample. For each parameter, the median value $\mu$~$\pm$~$c_{95}$ is reported, expressed in minutes. These offsets allow to better understand how the flares behave in the HXR and SXR regimes. The distributions show a skewness on the left-hand side, pointing out that the \textit{peak flux} in the SXR tends to occur earlier than in the HXR.}
\label{distribution_temporal_class}
\end{figure}

\begin{table}
\centering
\caption{Offset of the Flare Maxima between AGILE and GOES.}
\label{table_offsets}
\begin{center}
\begin{tabular}{|c|c|c|c|}
\hline
\textbf{Class} & $t_{max}^{AGILE}-t_{max}^{GOES}$ & $t_{max}^{AGILE}-t_{max}^{GOES}$ & \textbf{$\mu-m$} \\
~ & \textbf{Median $\mu$} & \textbf{Mean $m$} & \textbf{[minutes]} \\
~ & \textbf{[minutes]} & \textbf{[minutes]} & ~ \\ \hline \hline
B & -1.0 & -1.5 & 0.5 \\
C & -1.0 & -1.9 & 0.8 \\
M & -1.4 & -10.7 & 9.3 \\ \hline
TOT & -1.0 & -1.9 & 0.9 \\ \hline
\end{tabular}
\end{center}
Median $\mu$ and mean $m$ values of temporal offsets $t_{max}^{AGILE}-t_{max}^{GOES}$, for each intensity classes and for the total sample. The last column reports the skewness of the distribution, where positive values indicate that the flare \textit{maximum} is reached earlier in GOES, with respect to AGILE.
\end{table}

Fig.~\ref{distribution_temporal_class} shows the time differences of the \textit{maximum} times, as seen in the AGILE and in the GOES data, for the same detected solar flares. The distributions are divided into B, C, and M classes; in the last row, the distribution is evaluated on the total population. The $t_{max}^{AGILE}-t_{max}^{GOES}$ distributions show a rather steep profile, with a median value of about $\sim -1$~minute. The distributions show a general skewness on the left-hand side, suggesting that the \textit{peak flux} in the SXR is generally reached before the \textit{peak flux} in the HXR. This skewness can be quantified by studying the difference between the median $\mu$ and the mean $m$ values of each distribution: these differences $\mu-m$ turn out to be all greater than zero, implying that the mean value is located on the left of the median. This is particularly evident in the C class events ($\mu-m=0.8$~minutes), which exhibit the highest statistics, but a similar behavior could be hinted also in the other classes ($\mu-m=0.5$~min for class B and $\mu-m=2.7$~minutes for class M, affected by lower statistics). The values are reported in Tab.~\ref{table_offsets}.

For the sake of completeness, we evaluated whether such temporal differences could be ascribed to the different orbits of the satellites. A solar flare emitted at the Sun's surface would reach AGILE and GOES at slightly different times: however, in the worst case, such delay would translate into about $\sim0.1$~s, which is well below the AC-Lat4 1.024~s time resolution. The effects due to the photon time of propagation can therefore be neglected and the AGILE-GOES offset can be entirely ascribed to the different energy ranges of the two satellites.


\subsection{Solar Flare Intensity Parameters}
\label{solar_flare_peak_flux_and_fluence}

The AC-Lat4 is a plastic scintillation detector, collecting all counts within a single 80-200~keV energy channel, and integrating all of them in fixed 1.024~s bins. As a consequence, for the AGILE solar flares, no spectral analysis can be performed. However, it is possible to investigate the \textit{peak counts} reached by each event (proportional to the event \textit{peak flux}), as well as to evaluate the \textit{integrated counts} released by each flare within its \textit{duration} (proportional to the event \textit{fluence}). For this analysis, we considered the not saturated \textit{level 3} events. The \textit{peak counts} and \textit{integrated counts} are evaluated after the subtraction of the associated background count rate: the background is evaluated as the average count rate in the intervals from $t_{start}-30$~s to $t_{start}$ and from $t_{end}$ and $t_{end}+30$~s.

\begin{figure*}
\centering
\includegraphics[width=1.0\linewidth]{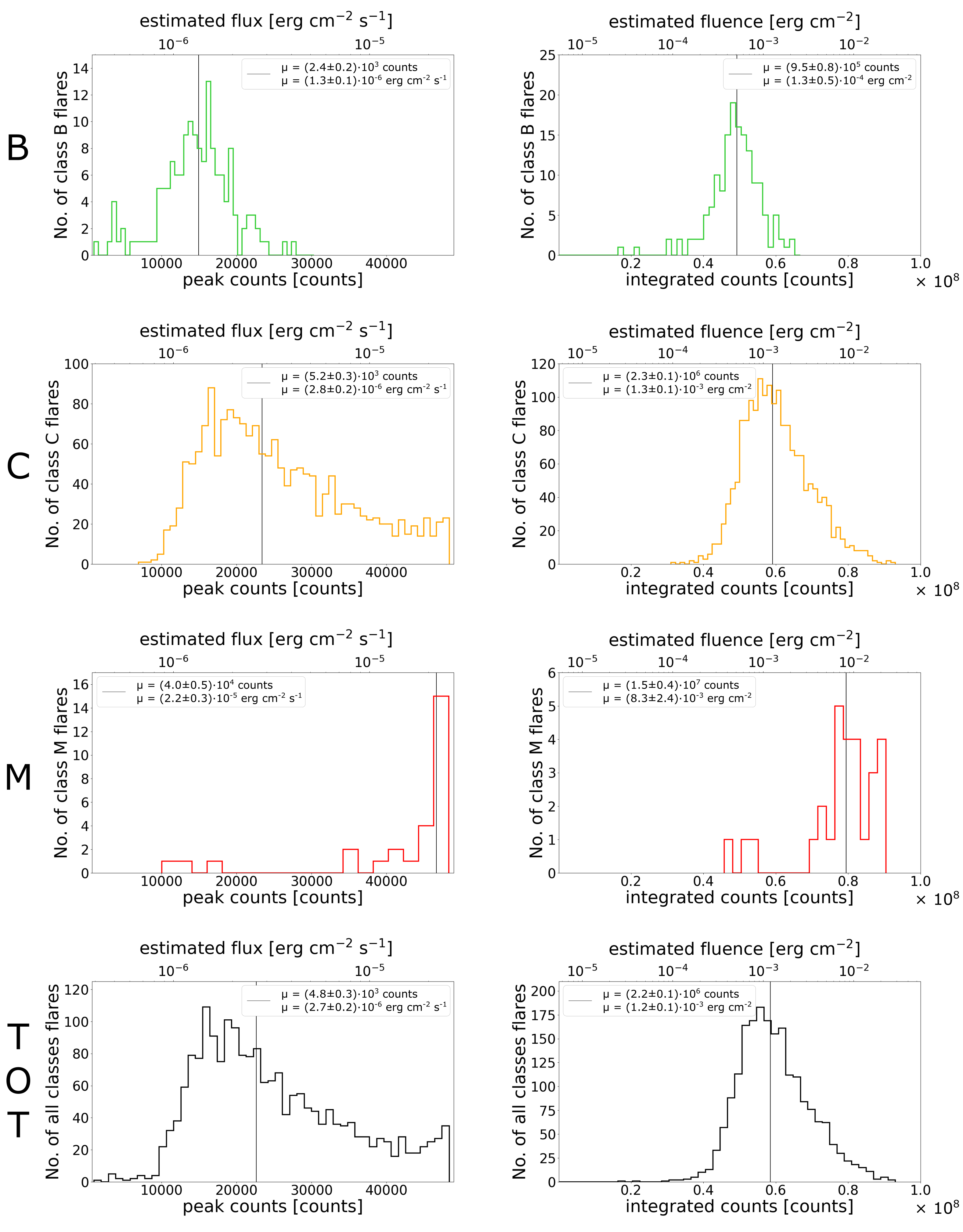}
\caption{Histograms of the \textit{peak counts} and \textit{integrated counts}, for the AGILE solar flares of \textit{level 3} sample. The first three rows refer to B, C, and M class events, while the last row includes the whole flare sample. Median values $\mu$ are reported as vertical lines. For each distribution, both the \textit{peak counts} and \textit{integrated counts} (bottom $x$-axes), as well as the corresponding estimated \textit{peak flux} and \textit{fluence} values in physical units (top $x$-axes) obtained from the calibration process, are reported.}
\label{distribution_intensity_class}
\end{figure*}

\begin{table}
\centering
\caption{Intensity Parameters of the AGILE \textit{level 3} Solar Flares}
\label{table_intensity_class}
\begin{center}
\begin{tabular}{|c|c|c|}
\hline
~ & \multicolumn{2}{|c|}{\textbf{Level 3}} \\ \hline
\textbf{Class} & \textbf{Parameter} & \textbf{Value} \\ \hline
\multirow{4}*{B} & peak counts & $(2.4\pm0.2)\times10^{3}$ \\
                 & peak flux [erg~cm$^{-2}$~s$^{-1}$] & $(1.3\pm0.1)\times10^{-6}$ \\ \cline{2-3}
                 & integrated counts & $(9.5\pm0.8)\times10^{5}$ \\
                 & fluence [erg~cm$^{-2}$] & $(5.2\pm0.5)\times10^{-4}$ \\ \hline
\multirow{4}*{C} & peak counts & $(5.2\pm0.3)\times10^{3}$ \\
                 & peak flux [erg~cm$^{-2}$~s$^{-1}$] & $(2.8\pm0.2)\times10^{-6}$ \\ \cline{2-3}
                 & integrated counts & $(2.3\pm0.1)\times10^{6}$ \\
                 & fluence [erg~cm$^{-2}$] & $(1.3\pm0.1)\times10^{-3}$ \\ \hline
\multirow{4}*{M} & peak counts & $(4.0\pm0.5)\times10^{4}$ \\
                 & peak flux [erg~cm$^{-2}$~s$^{-1}$] & $(2.2\pm0.3)\times10^{-5}$ \\ \cline{2-3}
                 & integrated counts & $(1.5\pm0.4)\times10^{7}$ \\
                 & fluence [erg~cm$^{-2}$] & $(8.3\pm2.4)\times10^{-3}$ \\ \hline \hline
\multirow{4}*{TOT} & peak counts & $(4.8\pm0.3)\times10^{3}$ \\
                   & peak flux [erg~cm$^{-2}$~s$^{-1}$] & $(2.7\pm0.2)\times10^{-6}$ \\ \cline{2-3}
                   & integrated counts & $(2.2\pm0.1)\times10^{6}$ \\
                   & fluence [erg~cm$^{-2}$] & $(1.2\pm0.1)\times10^{-3}$ \\ \hline
\end{tabular}
\end{center}
Median values $\mu$~$\pm$~$c_{95}$ of the \textit{peak counts} and \textit{integrated counts} distributions, and the corresponding \textit{peak flux} and \textit{fluence} distributions obtained with a preliminary calibration process described in Section~\ref{passingtophysicalunits}. Parameters are evaluated both on each intensity class and on the total sample.
\end{table}

Fig.~\ref{distribution_intensity_class} reports the \textit{peak counts} and the \textit{integrated counts} distributions, for the 1983 AGILE \textit{level 3} sample, divided into intensity classes. The intensity parameters are reported in both counts (bottom $x$-axes) and physical units (top $x$-axes), obtained from a preliminary calibration process that will be described in Section~\ref{passingtophysicalunits}. The \textit{peak counts} tend to increase with increasing the intensity, as expected from the definition of solar classes. Similarly, the median value of the \textit{integrated counts} distribution tends to assume larger values, with increasing flare class: this behavior is also expected, as high-intensity class flares show longer \textit{durations}, increasing the number of \textit{integrated counts}. The median values $\mu$ and the corresponding $c_{95}$ errors for the distributions of \textit{peak counts} and \textit{integrated counts} encountered in the different classes are reported in Tab.~\ref{table_intensity_class}, together with the corresponding \textit{peak fluxes} and \textit{fluences} obtained with the calibration process.


\subsection{Passing to Physical Units}
\label{passingtophysicalunits}

In order to retrieve the \textit{peak flux} and \textit{fluence}, we must convert the AC \textit{peak counts} and \textit{integrated counts} to physical units. This conversion would require the effective area of the detector, providing the sensitivity of the instrument at different energies under different incoming angles. However, such product is not available, as the AC system was not intended to perform detection of events, nor spectroscopic analysis, but only to act as the veto for background charged particles. As a consequence, in order to calibrate the detector, we use external events (i.e., GRBs), detected either by the AC-Lat4 or by other instruments on board other satellites (e.g., Konus-Wind, Fermi Gamma-Ray Burst Monitor, hereafter GBM). Using the \textit{fluence} of these bursts reported by the external measurements, integrating it in the AC energy range, and comparing it with the corresponding AC-Lat4 \textit{integrated counts}, we could obtain a preliminary calibration coefficient to pass from counts to physical units.

\begin{table}
\centering
\caption{Long GRBs Used as Calibrators}
\label{bursts}
\begin{center}
\begin{tabular}{|c|c|c|}
\hline
\textbf{Burst} & \textbf{Off Axis [deg]} & \textbf{Calibration Factor} \\ \hline
GRB 190415A & 0.72 & 4.1E-10 \\ \hline
GRB 160720A & 11.09 & 8.4E-10 \\ \hline
GRB 190727B & 12.58 & 5.5E-10 \\ \hline
GRB 171227A & 13.09 & 3.9E-10 \\ \hline
\end{tabular}
\end{center}
Four long GRBs detected by the AC-Lat4 with off axis angles $<15^{\circ}$. Second and third columns report the values of the off axis angles and calibration factors. The selected events exhibit an average calibration factor of $5.5\times10^{-10}$, which is used for converting the detector counts into physical units.
\end{table}


We carried out a search for GRBs in the AC-Lat4 database, by cross-correlating the AC data with the comprehensive Icecube GRBWeb catalog by P.~Coppin\footnote{https://user-web.icecube.wisc.edu/$\sim$grbweb\_public/}, already adopted for AGILE studies on GRBs \citep{Ursi2022b,Parmiggiani2023}. The cross-search ended up with 145 bursts detected by the AC-Lat4. In particular, 7 GRBs took place during ongoing solar flare emissions, pointing out the high occurrence of the solar events in the AC-Lat4 data. In order to deal with events with a sufficiently large number of counts and a reliable statistics, out of this sample we selected only the long GRBs, with a well-defined temporal structure. Furthermore, we took only those bursts exhibiting a stable background (i.e., not occurring during solar flares, nor near SAA regions), ending up with 26 bursts. For each GRB, we retrieved the best-fit spectral parameters provided by Konus-Wind or Fermi-GBM, and integrated the overall \textit{fluences} only in the 80-200~keV energy range of the AC-Lat4. We compared the obtained \textit{fluences} in the AC-Lat4 energy band to the AC-Lat4 \textit{integrated counts}, evaluated in the same time interval, ending up with a calibration factor for each event. For these 26 bursts, the calibration factor tends to assume slightly larger values with increasing the off axis angle. In particular, the distribution of the calibration factors with respect to the off axis angles is best-fitted with a power law function with index 0.3. As the AC-Lat4 is parallel to the solar panels and as the satellite is spinning about its sun-pointing axis, the entire solar flare sample exhibits small off axis angles, peaking at $4^{\circ}$, and extended up to at most $\sim15^{\circ}$ in some rare cases. We found 4 GRBs detected under angular configurations similar to those of solar flares (i.e., under off axis angles $<15^{\circ}$). These bursts are reported in Tab.~\ref{bursts}, with information about the off axis angle under which they have been detected, and the corresponding calibration factors. For these 4 events, the calibration factor assumes an average value of $m\sim5.5\times10^{-10}$. This value can be expressed as $m=F/N$, where $F$ is the \textit{fluence} and $N$ is the corresponding number of \textit{integrated counts}, in the same time interval. Expressing the \textit{fluence} as $F = K_{keV2erg} \cdot E_{avg} / A_{eff}(@E_{avg})$, where $K_{keV2erg}=1.6\times10^{-9}$~erg/keV is the conversion factor from keV to erg, $E_{avg} = 140$~keV is the average energy of the AC-Lat4 detector unique channel, we end up with an effective area of $A_{eff}(@E_{avg})\sim410$~cm$^{2}$. This value seems reasonable, considering that the AC-Lat4 detector has a geometrical surface of $\sim2200$~cm$^{2}$ \citep{Perotti2006}.

\begin{figure}
\centering
\includegraphics[width=1.0\linewidth]{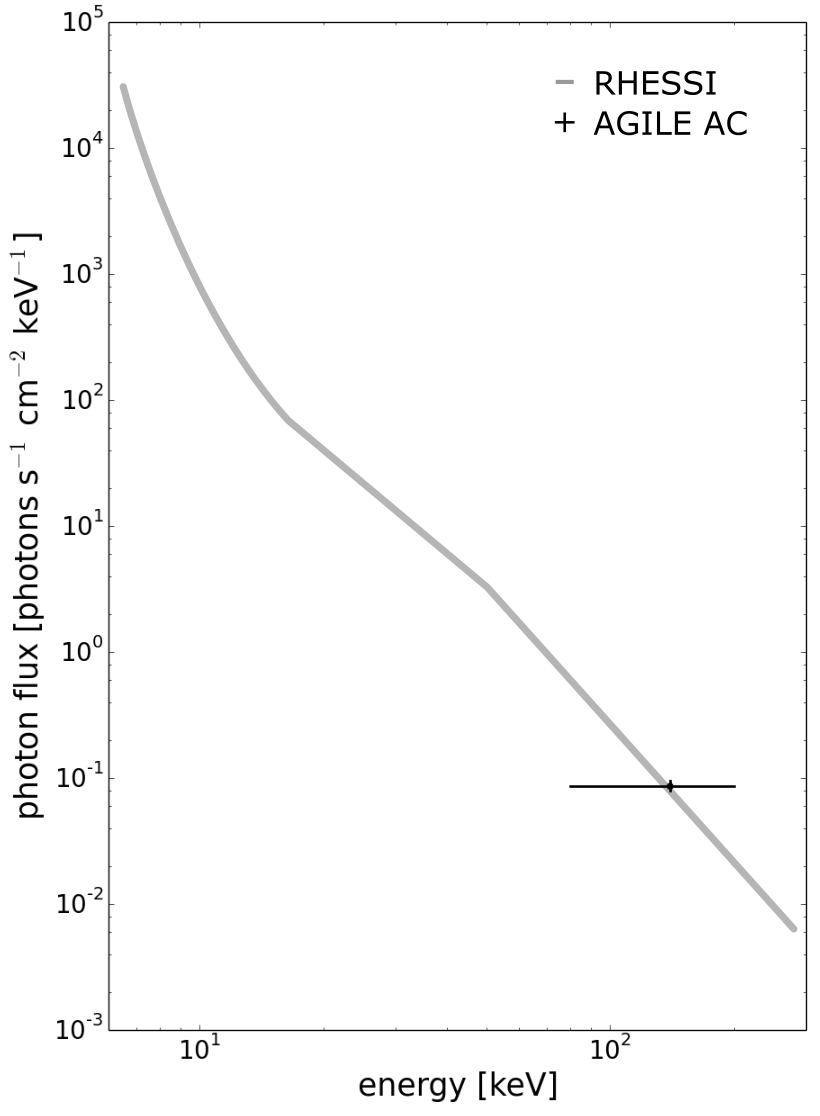}
\caption{Energy spectrum of a typical solar flare, extrapolated from \cite{Grigis&Benz2004,Benz2008} (gray), and the average flux obtained for the total AGILE solar flare sample, in the AC-Lat4 80-200~keV unique energy channel (black).}
\label{rhessi_spectrum}
\end{figure}

In Fig.~\ref{rhessi_spectrum}, we compared the average flux obtained for the AGILE total sample (i.e., the sample including all confirmed AGILE solar flares, regardless their intensity class) to a typical solar flare energy spectrum provided by RHESSI (for an event occurred on 2002 November 9 at 13:14:16 UT, reported by \cite{Grigis&Benz2004,Benz2008}). Although not allowing to carry out spectral analysis, the unique AGILE AC-Lat4 energy channel shows a reasonable agreement of the average flux value encountered by AGILE with the overall shape of a typical RHESSI flare spectrum. This points out that, although providing an approximation solution, our method allows a preliminary conversion of energy from the AC-Lat4 counts to physical units. For the sake of completeness, in Tab.~\ref{table_intensity_class} and Figs.~\ref{distribution_intensity_class},\ref{class_vs_intensity_ALL}, we present the intensity parameters expressed both in AC counts and in estimated physical units.

\subsection{Cross-correlation between AGILE and GOES}

\begin{figure*}
\centering
\includegraphics[width=1.0\linewidth]{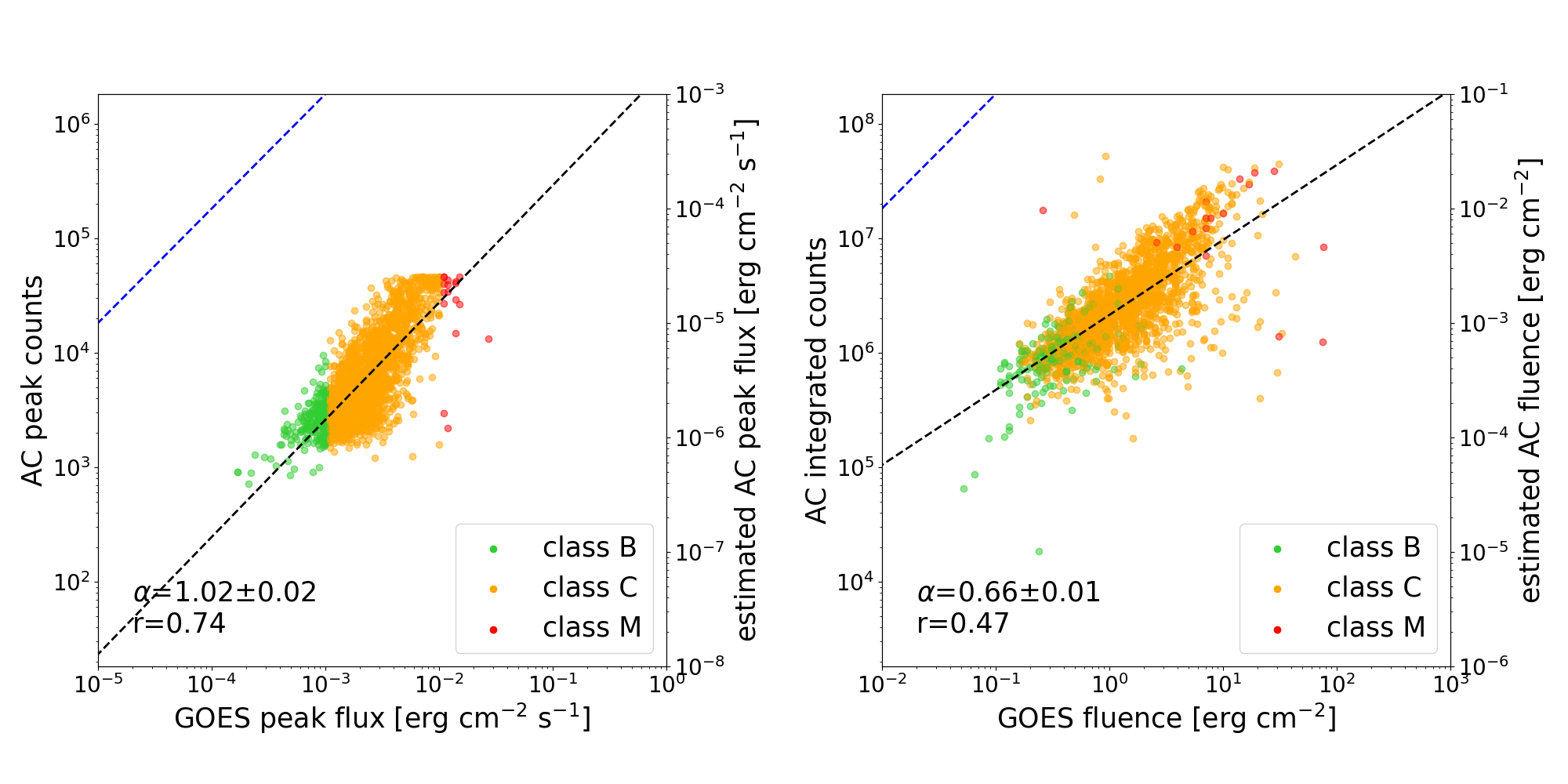}
\caption{Scatter plots of \textit{level 3} solar flares intensity parameters, as revealed in the GOES data (in the SXR range) with respect to the AGILE AC-Lat4 data (in the HXR range). The AGILE estimated \textit{peak flux} and \textit{fluence} parameters, labeled on the right \textit{y}-axes, are evaluated using the calibration process illustrated in Section~\ref{passingtophysicalunits}. The blue dashed lines represent the bisectors of the two plots, where points would be distributed if the parameters exhibited the same values in both energy ranges. The black dashed lines are the best fit obtained with power law models. The corresponding power law indices and cross-correlation factors are reported in the bottom left legends.}
\label{class_vs_intensity_ALL}
\end{figure*}

We cross-correlated the \textit{peak flux} and \textit{fluence} of the \textit{level 3} solar flares, as revealed in the GOES data (SXR regime) and in the AGILE data (HXR regime). The left plot of Fig.~\ref{class_vs_intensity_ALL} shows the scatter plot of the \textit{peak counts} (left $y$-axis) and estimated \textit{peak flux} (right $y$-axis) with respect to GOES \textit{peak flux}. The right plot shows the scatter plot of the \textit{integrated counts} (left $y$-axis) and estimated \textit{fluence} (right $y$-axis), with respect to GOES \textit{fluence}, obtained for the same sample of events. Flares belonging to different classes are represented in different colors. We report the corresponding bisectors of the two plots as blue dashed lines: such lines would represent the case in which the \textit{peak flux} and \textit{fluence} had the same values in both energy ranges, respectively. In the plots, the best-fit power laws are represented with black dashed lines. We notice that either the \textit{peak flux} or the \textit{fluence} provided by the GOES exhibit on average larger values with respect to those obtained with AGILE: this is in agreement with what found by \cite{Hudson1994}, who compared the SXR and HXR spectra, reporting that the emission in the SXR is usually brighter than a simple power law extrapolation from HXR down to the softer energy range, given the existence of both thermal and nonthermal components in the emission mechanism. The \textit{peak flux} scatter plot shows good cross-correlation, with a coefficient of $r=0.74$ and a best-fit power law with index $\alpha=1.02$. On the other hand, the \textit{fluence} scatter plot exhibits a more spread distribution, ending up with a lower cross-correlation coefficient of $r=0.47$ and a best-fit power law index of $\alpha=0.66$. This could be ascribed to the different definition of \textit{start} and \textit{end} times of the two satellites, which may result in underestimations or overestimates of the flare \textit{durations} and, consequently, of the associated \textit{fluence}. Although a univocal correlation between the GOES and the AGILE values is not achieved, given the relatively large span of both distributions, the best fit lines allow to discriminate among different regions of the parameter space, to identify lower- and higher-class events, on the basis of their \textit{peak counts} and \textit{integrated counts} revealed by AGILE. This could provide further information on the flares detected by the AC-Lat4, which can be inserted in the real-time analysis pipeline.


\section{Cross-check with Other High-energy Missions}

In order to carry out a comparison of the results obtained by the AGILE AC-Lat4 in the 80-200~keV energy range, we investigated the solar flares reported by other space missions operating in the high-energy regime. We retrieved and analyzed the public lists of solar flares reported by RHESSI and Fermi GBM \footnote{available at https://data.nas.nasa.gov/helio/portals/solarflares}. The RHESSI events are reported in eight different energy channels, covering from 3~keV to 7~MeV, whereas the Fermi GBM events are observed in the 8~keV-40~MeV energy range. At the time of this analysis, the RHESSI events covered a time interval from 2002 February 12 to 2018 March 30 and the Fermi GBM sample covers from 2008 November 2 to 2022 April 30.

The public RHESSI and Fermi GBM lists provide the times of the \textit{start}, \textit{end}, and \textit{maximum} of each detected flare, allowing to calculate the \textit{duration}, \textit{rise} and \textit{decay} time parameters for their samples; however, their lists only report intensity parameters expressed in counts, and not in physical units. As a consequence, the RHESSI and Fermi GBM flares have been exploited to cross-check: (1) the fraction of events detected in each intensity class; (2) the asymmetry between \textit{rise} and \textit{decay} phases; (3) the temporal offsets of flare maxima with respect to GOES. Similarly to what carried out for AGILE, the \textit{duration} has been computed as the difference between the flare \textit{start} and \textit{end} times, whereas the \textit{rise} and \textit{decay} times were evaluated as the differences between the flare \textit{maximum} and the \textit{start} time, and between the \textit{end} time and the \textit{maximum}, respectively.

As shown in the following subsections, these cross-checks provided good confirmations of the temporal features found from the AGILE solar flares in the high-energy regime.

\subsection{Fraction of Flares per Class}

Out of the RHESSI and Fermi GBM lists, we selected all events with a clear GOES association, similarly to what carried out for the AGILE sample. We ended up with different fractions of solar flares, revealed in different intensity classes. These values are reported in Tab.~\ref{classes_missions}. We notice that the percentages obtained from the Fermi GBM dataset seem the closest ones to those provided by AGILE, given the larger similarity between the two satellites, with C class events ranging within 81\%-82\% and M class events within 13\%-15\%. On the other hand, the RHESSI sample is the most similar to GOES for what concerns the amount of B class flares, constituting a nonnegligible fraction of the overall sample.

\begin{table}
\centering
\caption{Solar Flares Detected by AGILE Compared with Those Reported by Other Space Missions.}
\label{classes_missions}
\begin{center}
\begin{tabular}{|c|c|c|}
\hline
\textbf{Mission} & \textbf{Class} & \textbf{Fraction of Events} \\ \hline \hline
~                    & A & 0.5\% \\
~                    & B & 47.6\% \\
GOES                 & C & 47.2\% \\
(1.5-12.5~keV)       & M & 4.4\% \\
~                    & X & 0.3\% \\ \hline
~                    & A & 0.1\% \\
~                    & B & 39.7\% \\
RHESSI               & C & 54.5\% \\
(3~keV-7~MeV)        & M & 5.3\% \\
~                    & X & 0.4\% \\ \hline
~                    & A & 0\% \\
~                    & B & 4.9\% \\
AGILE                & C & 81.3\% \\
(80-200~keV)         & M & 13.0\% \\
~                    & X & 0.8\% \\ \hline \hline
~                    & A & 0\% \\
~                    & B & 1.9\% \\
Fermi GBM            & C & 82.0\% \\
(8~keV-40~MeV)       & M & 15.0\% \\
~                    & X & 1.1\% \\ \hline
\end{tabular}
\end{center}
Fractions of AGILE, GOES, Fermi GBM, and RHESSI solar flares, divided into intensity classes. The different percentages of events revealed in each class depend on the energy ranges and sensitivities of the satellites.
\end{table}

\subsection{\textit{Decay/Rise} Ratio}
\label{decayriseratio}

\begin{table*}
\centering
\caption{\textit{Duration}, \textit{Rise} Time, and \textit{Decay} Time of the AGILE Flares Compared with Other Space Missions}
\label{table_highenergyduration}
\begin{center}
\begin{tabular}{|c|c|c|c|c|c|}
\hline
\textbf{Mission} & \textbf{Class} & \textbf{Duration} & \textbf{Rise Time} & \textbf{Decay Time}& \textbf{Decay/} \\
\textbf{~} & ~ & \textbf{[minutes]} & \textbf{[minutes]} & \textbf{[minutes]} & \textbf{Rise} \\ \hline \hline
~                     & A & $3.0\pm1.0$ & $1.0\pm0.5$ & $1.0\pm0.3$ & 1.0 \\
~                     & B & $12.0\pm0.2$ & $6.0\pm0.1$ & $5.0\pm0.1$ & 0.8 \\
GOES                  & C & $14.0\pm0.2$ & $7.0\pm0.1$ & $7.0\pm0.1$ & 1.0 \\
(1.5-12.5~keV)        & M & $19.0\pm1.1$ & $10.0\pm0.6$ & $8.0\pm0.5$ & 0.8 \\
~                     & X & $23.0\pm5.9$ & $16.0\pm2.9$ & $9.0\pm2.7$ & 0.6 \\ \cline{2-6}
~                     & TOT & $13.0\pm0.1$ & $7.0\pm0.1$ & $6.0\pm0.1$ & 0.9 \\ \hline
~                     & A & $3.6\pm1.2$ & $1.4\pm0.2$ & $2.2\pm1.0$ & 1.6 \\
~                     & B & $9.2\pm0.2$ & $2.8\pm0.1$ & $6.0\pm0.2$ & 2.2 \\
RHESSI                & C & $16.3\pm0.3$ & $5.0\pm0.1$ & $10.2\pm0.2$ & 2.0 \\
(3~keV-7~MeV)         & M & $24.9\pm1.5$ & $7.4\pm0.6$ & $15.7\pm1.2$ & 2.1 \\
~                     & X & $26.5\pm5.3$ & $9.7\pm1.7$ & $17.6\pm3.7$ & 1.8 \\ \cline{2-6}
~                     & TOT & $13.3\pm0.2$ & $4.0\pm0.1$ & $8.4\pm0.2$ & 2.1 \\ \hline
~                     & A & --- & --- & --- & --- \\
~                     & B & $10.1\pm0.8$ & $4.1\pm0.5$ & $5.6\pm0.6$ & 1.4\\
AGILE                 & C & $13.0\pm0.3$ & $5.0\pm0.2$ & $7.4\pm0.2$ & 1.5 \\
(80-200~keV)          & M & $20.4\pm1.9$ & $6.8\pm2.5$ & $12.4\pm2.8$ & 1.8 \\
~                     & X & $32.8\pm1.7$ & --- & --- & --- \\ \cline{2-6}
~                     & TOT & $13.1\pm0.3$ & $4.9\pm0.2$ & $7.6\pm0.2$ & 1.6 \\ \hline \hline
~                     & A & --- & --- & --- & --- \\
~                     & B & $1.7\pm0.3$ & $0.5\pm0.1$ & $1.1\pm0.3$ & 2.2 \\
Fermi GBM             & C & $6.4\pm0.3$ & $1.4\pm0.1$ & $4.4\pm0.2$ & 3.2 \\
(8~keV-40~MeV)        & M & $24.4\pm1.9$ & $5.2\pm0.5$ & $16.7\pm1.4$ & 3.2 \\
~                     & X & $40.6\pm8.6$ & $11.2\pm4.6$ & $22.1\pm4.8$ & 2.0 \\ \cline{2-6}
~                     & TOT & $7.7\pm0.4$ & $1.6\pm0.1$ & $5.3\pm0.3$ & 3.2 \\ \hline
\end{tabular}
\end{center}
Median values $\mu$~$\pm$~$c_{95}$ of the \textit{duration}, \textit{rise}, and \textit{decay} times obtained for the AGILE solar flares, compared to GOES, RHESSI, and Fermi GBM events. Data refer both to the different intensity classes and the total sample. The last column reports the ratio between the \textit{decay} and the \textit{rise} time, providing a good estimate of the asymmetry between the two phases. Such asymmetry is evident in the AGILE, RHESSI, and Fermi GBM datasets, acquired in the HXR range, while the GOES events seem to exhibit rather symmetric \textit{rise} and \textit{decay} phases.
\end{table*}

We investigated the asymmetry between \textit{rise} and \textit{decay} times, previously obtained for the AGILE solar flares. From the RHESSI and Fermi GBM subsamples of flares with clear GOES association, we evaluated the \textit{duration}, \textit{rise}, and \textit{decay} time distributions. We report in Tab.~\ref{table_highenergyduration} these parameters, for the various space missions. In the last column, we report the ratio between the \textit{decay} and the \textit{rise} time, which provides an indication of how much longer does the \textit{decay} phase lasts with respect to the \textit{rise} phase. This parameter is useful, as it provides an estimate of the asymmetry between \textit{rise} and \textit{decay} times for each satellite. It can be noticed that a slightly more pronounced asymmetry between \textit{rise} and \textit{decay} times is present in the samples acquired by the high-energy missions. In particular, while \textit{rise} and \textit{decay} times in GOES look very similar for each intensity class, with the total sample showing a \textit{decay/rise}~=~0.9, the AGILE, RHESSI, and Fermi GBM \textit{decay} times are 1.4-3.2 times greater than their respective \textit{rise} times: this means that the \textit{decay} phases constitute 60\%-75\% of the entire flare \textit{duration} in the high-energy data. Such asymmetry is evident in the AGILE, RHESSI, and Fermi GBM detected events, but it does not exhibit a clear dependence on the associated intensity class. This longer persistence of the flares in the HXR regime with respect to the SXR regime could represent a hint that more intense, harder-spectrum solar flares could exhibit longer-lasting \textit{decay} phases. Also, such evidence of an extended HXR component, following a former impulsive SXR flare, could represent a hint of the two-phase acceleration process \citep{McTiernan1993}, with the second step allowing particles to reach higher energies ($>100$~keV) \citep{Frost&Dennis1971}.

\begin{table*}
\centering
\caption{Offset of the Flare Maxima between AGILE, Fermi GBM, RHESSI, and GOES}
\label{table_offsets_missions}
\begin{center}
\begin{tabular}{|c|c|c|c|c|}
\hline
\textbf{Mission} & \textbf{Class} & $t_{max}^{mission}-t_{max}^{GOES}$ & $t_{max}^{mission}-t_{max}^{GOES}$ & \textbf{$\mu-m$} \\
~ & ~ & \textbf{Median $\mu$} & \textbf{Mean $m$} & \textbf{[minutes]} \\
~ & ~ & \textbf{[minutes]} & \textbf{[minutes]} & ~ \\ \hline \hline
~                    & A & --- & --- & --- \\
~                    & B & -1.0 & -1.5 & 0.5 \\
AGILE                & C & -1.0 & -1.9 & 0.8 \\
(80-200~keV)         & M & -1.4 & -10.7 & 9.3 \\ \cline{2-5}
~                    & TOT & -1.0 & -1.9 & 0.9 \\ \hline \hline
~                    & A & 0.1 & 0.3 & -0.1 \\
~                    & B & -0.7 & -1.2 & 0.5 \\
RHESSI               & C & -0.9 & -1.6 & 0.7 \\
(3~keV-7~MeV)        & M & -1.2 & -2.4 & 1.0 \\
~                    & X & -0.9 & -2.5 & 1.6 \\ \cline{2-5}
~                    & TOT & -0.8 & -1.5 & 0.6 \\ \hline
~                    & A & --- & --- & --- \\
~                    & B & -0.6 & -0.6 & 0.1 \\
Fermi GBM            & C & -1.4 & -2.3 & 0.9 \\
(8~keV-40~MeV)       & M & -1.8 & -3.1 & 1.3 \\
~                    & X & -0.8 & -1.5 & 0.7 \\ \cline{2-5}
~                    & TOT & -1.4 & -2.3 & 0.9 \\ \hline
\end{tabular}
\end{center}
Median $\mu$ and mean $m$ values of temporal offsets $t_{max}^{mission}-t_{max}^{GOES}$, for AGILE, Fermi GBM, and RHESSI missions, evaluated with respect to GOES. For each satellite, we consider different intensity classes and the total samples. The last column measures the skewness of the distribution, where positive values indicate a bias of the distribution to the left, pointing out that the flare \textit{maximum} is reached later in the high-energy satellites data with respect to GOES.
\end{table*}

\subsection{Temporal Offset with Respect to GOES}
\label{temporal_offset_with_respect_to_goes}

We also exploited the RHESSI and Fermi GBM data to cross-check the offset of the \textit{maximum} with respect to GOES, previously found in the AGILE events. Evidence of an offset with respect to the GOES \textit{maximum} time can be noticed also in their datasets, as shown in Tab.~\ref{table_offsets_missions}. We noticed that, except for the low-populated RHESSI A class, all other classes of AGILE, Fermi GBM, and RHESSI show $\mu-m>0$, pointing out that the flare maxima in the high-energy data tend to be observed later, with respect to GOES. Their offset distributions show skewnesses which seem to get more pronounced with increasing flare class. This is particularly evident in the RHESSI data. On the other hand, the behavior of the Fermi X class and the AGILE M class events may not be reliable, due to the low number statistics.

From our analysis, the peak in the HXR seems to occur on average between 1.5~minutes (for RHESSI) and 2.3~minutes (for Fermi GBM) after that in the SXR. Although there have been cases of very intense solar flares, whose maxima in different wavelengths were detected within $\pm1$~s \citep{Kane1986}, the existence of a minute-scale offset in the occurrence of the peaks in SXR and HXR represents a further hint that the acceleration process is articulated into two steps, similarly to what we pointed out in Section~\ref{decayriseratio}.


\section{Other Unmatched Transients Detected by AGILE}
\label{other_transients_detected_by_agile}

For completeness, we focused on the sample of 1424 AGILE unmatched transients. As pointed out in Section~\ref{data_levels}, these transients were identified by the same algorithm that identified the matched solar flares, but the cross-search with the GOES data did not result in any match. However, as it will be illustrated in the following subsections, a large fraction of them exhibits close temporal correspondences in other solar flare lists, reported by other space missions such as Fermi GBM and RHESSI. Adopting the same categories used to classify the matched solar flare sample into different levels, we find out that, out of these 1424 events (\textit{level n1}), 998 have been fully acquired on board with no interruptions due to SAA and EOs (\textit{level n2}). Moreover, out of them, 994 did not saturate the on board detector (\textit{level n3}).

\begin{table}
\centering
\caption{AGILE \textit{Level 1-3} Matched Flares and \textit{Level n1-3} Unmatched Transients}
\label{nogoes}
\begin{center}
\begin{tabular}{|c|c|c|c|}
\hline
\textbf{Class}   & \textbf{Level 1} & \textbf{Level 2} & \textbf{Level 3}  \\
\textbf{(from}   & \textbf{(with GOES}    & \textbf{(Fully}      & \textbf{(Not}   \\
\textbf{GOES)}   & \textbf{Association)}  & \textbf{Acquired)}     & \textbf{Saturated)}   \\ \hline
B   & 174                 & 155                    & 155                  \\ \hline
C   & 2904                & 1848                   & 1800                 \\ \hline
M   & 466                 & 141                    & 28                   \\ \hline
X   & 28                  & 5                      & 0                    \\ \hline
TOT & 3572                & 2149                   & 1983                 \\ \hline \hline
~ & \textbf{Level n1} & \textbf{Level n2} & \textbf{Level n3} \\
~ & \textbf{(no GOES} & \textbf{(Fully} & \textbf{(Not} \\
~ & \textbf{Association)} & \textbf{Acquired)} & \textbf{Saturated)} \\ \hline
~ & 1424 & 998 & 994 \\ \hline
\end{tabular}
\end{center}
Number of \textit{level 1}, \textit{level 2}, and \textit{level 3} matched solar flares, divided into classes, compared to the \textit{level n1}, \textit{level n2}, and \textit{level n3} AGILE unmatched transients.
\end{table}

Tab.~\ref{nogoes} reports the different number of \textit{level 1}, \textit{level 2}, and \textit{level 3} matched solar flares already reported in Tab.~\ref{classes}, compared to the \textit{level n1}, \textit{level n2}, and \textit{level n3} AGILE unmatched transients, for which no association with external GOES detections was fulfilled. It can be noticed that the fraction of \textit{level n1} transients for which a cross-correlation with GOES data was not fulfilled consists in about 28\% of the total sample obtained by the search algorithm. Out of them, about 70\% of the events does not suffer EOs or SAA, constituting the \textit{level n2} dataset: this value is compatible with what is found for the \textit{level 2} matched sample (60\%). In particular, it lays within the fractions reported for flares of class-B (89\%) and class-C (64\%). Finally, the fraction of events that did not saturate the AC count rate, out of the \textit{level n2} transients, is equal to more than 99\%, consisting in the \textit{level n3} sample: also this value is compatible with the fraction of \textit{level 3} events out of the \textit{level 2} sample (92\%). In particular, also this value lays within the fractions reported for flares of class-B (100\%) and class-C (97\%). In the end, the analysis of these different fractions of events in each level points out that these transients mostly behave as solar flares of class-B and class-C.

\subsection{Unmatched Transients Occurrence Rate}

\begin{figure*}
\centering
\includegraphics[width=1.0\linewidth]{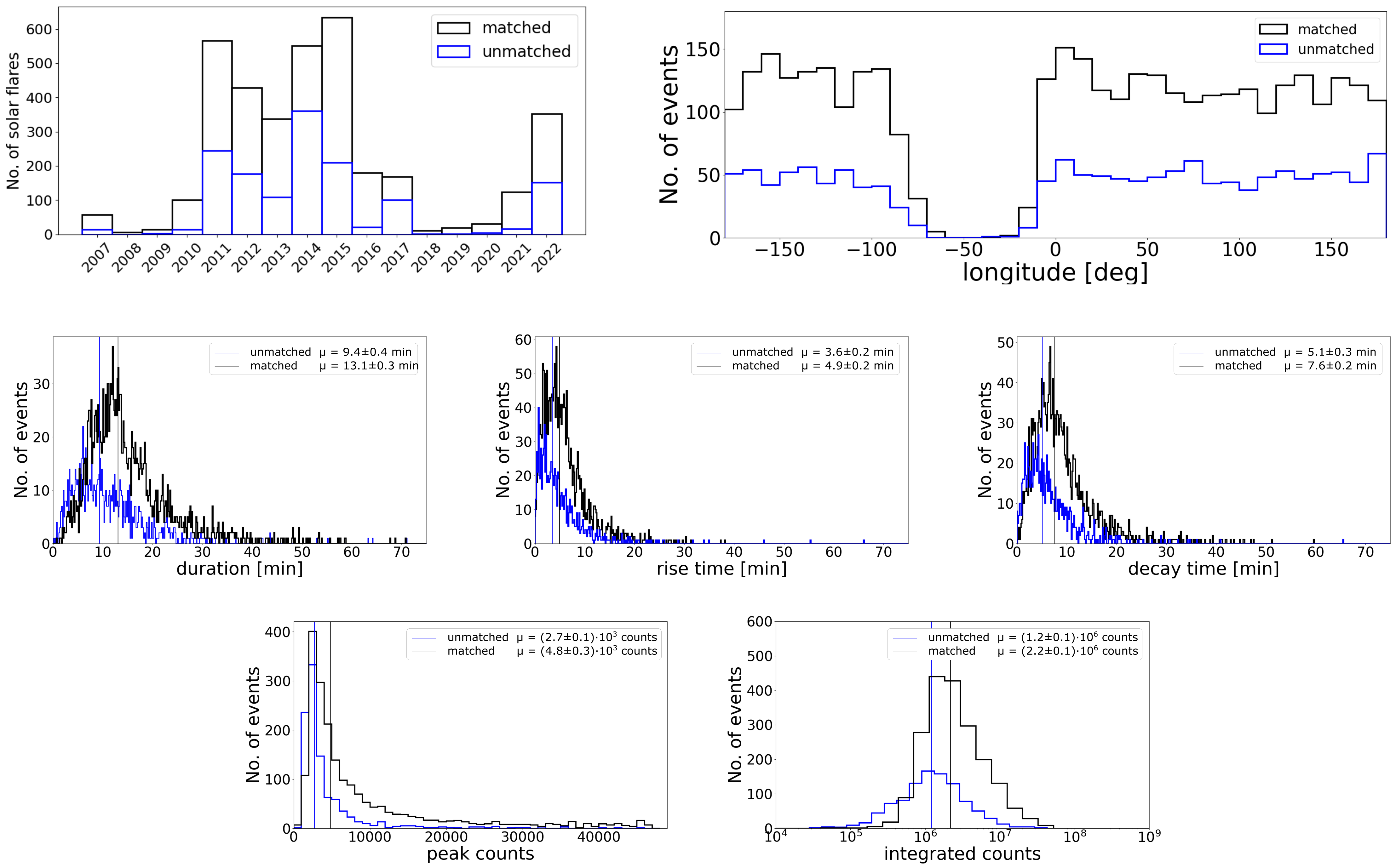}
\caption{Top left: rate of the AGILE 3572 matched solar flares (black) compared to the rate of the 1424 \textit{level n1} unmatched transients (blue). The general detection rates are very similar, exhibiting a p-value of 0.43 after a Kolmogorov-Smirnov test. Top right: longitude distribution of the AGILE subsatellite points at the time of the \textit{level 1} (black) and \textit{level n1} (blue) events. The lack of data corresponds to the South Atlantic Anomaly (SAA). Both distributions exhibit similar profiles, with no excesses along the orbit. Second row: distribution of \textit{duration}, \textit{rise} time, and \textit{decay} time of the 998 \textit{level n2} events (blue), together with the corresponding distributions evaluated for the 2149 \textit{level 2} solar flare sample (black). For each parameter, the median value $\mu$ is shown. Third row: distribution of \textit{peak counts} and \textit{integrated counts} of the 994 \textit{level n3} candidate transients fully detected by AGILE (blue), together with the corresponding distributions evaluated for the 1983 \textit{level 3} solar flare sample (black). For each parameter, median value $\mu$ is shown.}
\label{distribution_ncnc}
\end{figure*}

In order to investigate whether the unmatched transients clustered in time within a particular period, as a consequence of possible AGILE or GOES data losses, temporary failures, or worsening of their detection sensitivity, we evaluated the yearly occurrence rate of the unmatched transients, and compared it to the detection rate of the matched flares. The occurrence of \textit{level 1} transients and \textit{level n1} looks quite compatible, showing the same behavior throughout the years, both clearly following the solar cycle occurrence, as reported in the top left panel of Fig.~\ref{distribution_ncnc}. A Kolmogorov-Smirnov test of the compatibility of these two distributions, considering a confidence level of 95\%, results in a p-value of 0.43, which suggests that the two samples are compatible with the same yearly distribution.

\subsection{Subsatellite Point Geographic FDistribution}

We evaluated the geographic position of the satellite at each detection of unmatched transients, ending up with the distribution reported in the top right panel of Fig.~\ref{distribution_ncnc}. We performed this check to evaluate whether there exists a bias in the geographic distribution of the unmatched sample, and if it follows the same longitude distribution of the genuine solar flares. The two distributions show similar behaviors, with expected gaps near the SAA for both datasets, and not exhibiting clusterings throughout the orbit.

\subsection{Unmatched Transients Temporal and Intensity Parameters}

We studied the \textit{duration}, \textit{rise}, \textit{decay} times, \textit{peak counts}, and \textit{integrated counts} parameters of the AGILE unmatched transients, and compared those to the corresponding total distributions obtained for the matched solar flares. Similarly to what carried out for the matched sample, we adopt the fully acquired \textit{level n2} events to study the \textit{duration} of the transients and the \textit{level n3} events to evaluate \textit{rise} and \textit{decay} times, \textit{peak counts}, and \textit{integrated counts}.

\begin{table*}
\centering
\caption{Temporal and Intensity Parameters of the AGILE Matched Flares and Unmatched Transients}
\label{table_duration_ncnc}
\begin{center}
\begin{tabular}{|c|c|c|c|c|c|}
\hline
\textbf{Class} & \textbf{Level 1} & \multicolumn{2}{|c|}{\textbf{Level 2}} & \multicolumn{2}{|c|}{\textbf{Level 3}} \\ \hline
~ & \textbf{Duration} & \textbf{Rise Time} & \textbf{Decay Time} & \textbf{Peak Counts} & \textbf{Integrated Counts} \\
~ & \textbf{[minutes]} & \textbf{[minutes]} & \textbf{[minutes]} & ~ & ~ \\ \hline
B & $10.1\pm0.8$ & $4.1\pm0.5$ & $5.6\pm0.6$ & $(2.4\pm0.2)\times10^{3}$ & $(9.5\pm0.8)\times10^{5}$ \\
C & $13.0\pm0.3$ & $5.0\pm0.2$ & $7.4\pm0.2$ & $(5.2\pm0.3)\times10^{3}$ & $(2.3\pm0.1)\times10^{6}$ \\
M & $20.4\pm1.9$ & $6.8\pm2.5$ & $12.4\pm2.8$ & $(4.0\pm0.5)\times10^{4}$ & $(1.5\pm0.4)\times10^{7}$ \\
X & $32.8\pm1.7$ & --- & --- & --- & --- \\
TOT & $13.1\pm0.3$ & $4.9\pm0.2$ & $7.6\pm0.2$ & $(4.8\pm0.3)\times10^{3}$ & $(2.2\pm0.1)\times10^{6}$ \\ \hline \hline
~ & \textbf{Level n1} & \multicolumn{2}{|c|}{\textbf{Level n2}} & \multicolumn{2}{|c|}{\textbf{Level n3}} \\ \hline
~ & \textbf{Duration} & \textbf{Rise Time} & \textbf{Decay Time} & \textbf{Peak Counts} & \textbf{Integrated Counts} \\
~ & \textbf{[min]utes} & \textbf{[minutes]} & \textbf{[minutes]} & ~ & ~ \\ \hline
~ & $9.4\pm0.4$ & $3.6\pm0.2$ & $5.1\pm0.3$ & $(2.7\pm0.1)\times10^{3}$ & $(1.2\pm0.1)\times10^{6}$ \\ \hline
\end{tabular}
\end{center}
Median values $\mu$~$\pm$~$c_{95}$ of the \textit{duration}, \textit{rise} time, \textit{decay} time, \textit{peak counts}, and \textit{integrated counts} for the AGILE \textit{level 1-3} matched solar flares, compared to the \textit{level n1-3} unmatched transients.
\end{table*}

The second row of Fig.~\ref{distribution_ncnc} shows the \textit{duration}, \textit{rise} and \textit{decay} time distributions for the matched and unmatched samples, together with the corresponding median values, also reported in Tab.~\ref{table_duration_ncnc}. Although exhibiting slightly lower values with respect to the AGILE matched sample, the median \textit{duration}, \textit{rise} and \textit{decay} times are compatible with those of solar flares. The same asymmetry between \textit{rise} time and \textit{decay} time, encountered in the matched sample, is obtained also for the unmatched transients. In particular, the temporal parameters are mostly compatible with solar flares of class-B. The \textit{peak counts} and the \textit{integrated counts} distributions of the unmatched sample look similar to those of genuine solar flares, as reported in the third row of Fig.~\ref{distribution_ncnc}. The median values of these distributions are smaller than those of the entire \textit{level 3} population, reported in Tab.~\ref{table_duration_ncnc}. The \textit{peak counts} seem to be compatible with solar flares of class B, whereas the \textit{integrated counts} are more similar to values obtained in between B- and C class flares. In sum, the unmatched transients could be compatible with short-duration, low-intensity solar flares.

\subsection{Possible Explanations for the \textit{level n1} events}

We point out several possible reasons for which the unmatched AGILE transients are not reported as solar flares by the GOES satellites:

\begin{itemize}

\item \textit{GOES sensitivity and classification of solar flares.} Some of the nondetections of GOES satellites could be ascribed to their sensitivity to lower-intensity events. We know that the GOES classification into classes starts from class B1.0, which is the lowest intensity detectable by those satellites. Some of the \textit{level n1} AGILE events could be lower-intensity class-B (or class-A) solar flares, which have not been clearly distinguished and classified by the GOES detection algorithm. We investigated the temporal difference between the occurrence of the AGILE maxima (of both the matched and the unmatched events) and the GOES \textit{start} and \textit{end} times. Fig.~\ref{offset_all} shows the offset between AC-Lat4 \textit{maximum} and GOES \textit{start} and \textit{end} times, for both the \textit{level 2} solar flares and the \textit{level n2} transients. The matched sample shows a profile with AGILE maxima typically occurring after the GOES \textit{start} time and before the GOES \textit{end} time, as expected from a sample that has fulfilled the cross-search algorithm. On the other hand, the unmatched events exhibit quite different behaviors. The $t_{max}^{AGILE}-t_{start}^{GOES}$ difference shows a small excess before the zero, whereas the $t_{max}^{AGILE}-t_{min}^{GOES}$ delay shows an evident peak after the zero: these results show that a fraction of the unmatched events does not occur randomly, but tends to cluster nearby the GOES \textit{start-end} time intervals (i.e., just before the GOES \textit{start} time or just after the GOES \textit{end} time). For about $\sim150$ \textit{level n1} transients, the AGILE \textit{maximum} could occur some minutes after the GOES end. This behavior could be ascribed to the GOES classification method, for which when two or more solar flares occur close in time, the algorithm tends to classify only the most intense of them, neglecting the following lower-intensity one(s).

\begin{figure*}
\centering
\includegraphics[width=1.0\linewidth]{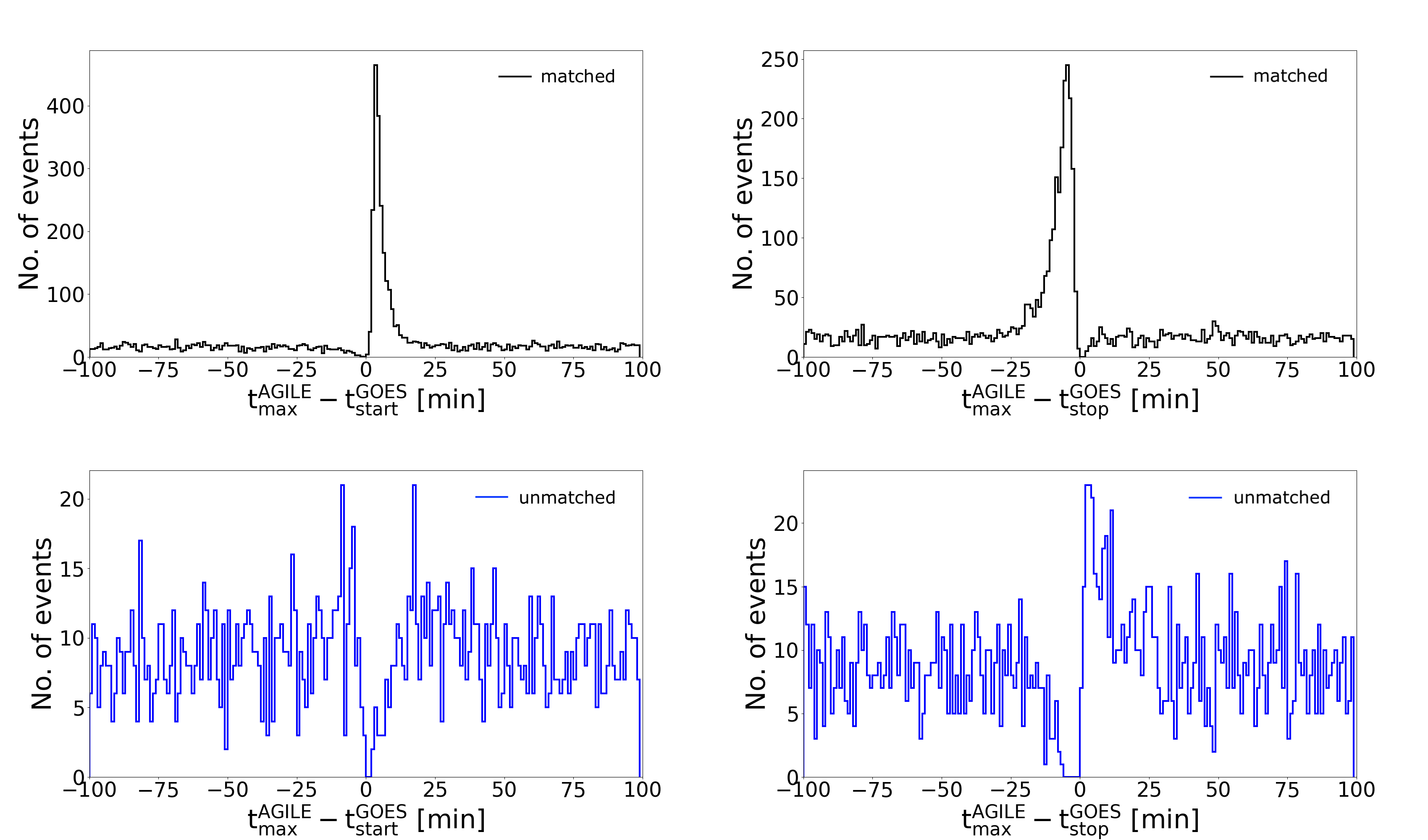}
\caption{Temporal offset between AC-Lat4 \textit{maximum} and GOES \textit{start} time (left) and GOES \textit{end} time (right), in the matched (black) \textit{level 2} and unmatched (blue) \textit{level n2} sample. For the matched sample, the AGILE \textit{maximum} tends to occur after the GOES \textit{start} time and before the GOES \textit{end} time, while for the unmatched sample, a large fraction of events seems to take place after the GOES \textit{end} time of the nearest solar flare.}
\label{offset_all}
\end{figure*}

\item \textit{GOES detection algorithm.} In a recent study, \cite{Plutino2022} identified a large fraction of new solar flares in the GOES data, by adopting a more effective detection algorithm. Their method allowed to expand the GOES solar flare database by about a factor of 5, in the period from 1986 to 2020, revealing a large sample of lower-intensity events. A cross-check between the AGILE \textit{level n1} transients and the newly identified GOES flares ended up with 553 matches, pointing out that a nonnegligible fraction of the unmatched events is actually solar flares, although not classified by the official GOES pipeline.

\item \textit{Cross-check with other missions.} We carried out a cross-check with the solar flare lists reported by RHESSI and Fermi GBM. The distribution of the time difference between the AGILE \textit{level n1} maxima and the Fermi GBM solar flare maxima peaks within $\pm2$~min, with 276 events occurring in that time window. Similarly, the distribution of the time difference between the AGILE \textit{level n1} maxima and the RHESSI solar flare maxima peaks within $\pm2$~minutes, with 470 events occurring in that time window. Out of them, 126 events are reported in both samples. Moreover, out of them, 85 are also present in the new GOES solar flare database. This means that $276+470+553-85=1214$ events of the \textit{level n1} dataset have been not reported in the official GOES dataset, but are reported in the Fermi GBM, RHESSI, and new GOES samples. These independent detections point out the genuine nature of the largest fraction (i.e., 85\%) of the \textit{level n1} AGILE transients, although not accounting for the entire 1424 events, leaving 210 unmatched transients. However, we recall that Fermi and RHESSI are LEO satellites as well, and their detections of solar flares are affected by EOs in the same measure as AGILE: this implies that a nonnegligible fraction (about a half, as pointed out in Section~\ref{issues}) of the \textit{level n1} events may have not been detected by these satellites. We also recall that the new GOES sample provided by \cite{Plutino2022} includes flares until 2020, not covering the last two yr of the AGILE analysis.

\item \textit{Earth occulations of GOES.} We investigated the possibility that these transients were detected by AGILE during EOs of the Sun for the GOES satellites. As explained in Section~\ref{issues}, in each orbit, a geostationary satellite suffers an EO of the Sun equal to about $5\%$ of its entire path. Assuming a homogeneous distribution of the AGILE detected flares, we end up with $5003\cdot0.05\sim250$ events which could have not been detected by GOES due to occultations. Such a number is totally compatible with the number of remaining unmatched transients (i.e., 210), for which no matches with Fermi, RHESSI, or new GOES solar flares are obtained.

\end{itemize}


\section{The AGILE Solar Flare Online Catalog}

The complete catalog of AGILE solar flares can be found on the ASI Space Science Data Center (SSDC) \textit{https://www.ssdc.asi.it/agilesolarcat/}. It reports the 5003 solar transients detected by the AGILE AC-Lat4, discussed in this work, with information about the different parameters of each detected event. The catalog includes different fields, containing information about the data level, the flare intensity class (if a GOES association is found), the \textit{start} time, the \textit{maximum} time, the \textit{end} time, the \textit{duration}, the \textit{rise} time, the \textit{decay} time, the $t_{max}^{AGILE}-t_{max}^{GOES}$ offset, the \textit{peak counts}, estimated \textit{peak flux}, the \textit{integrated counts}, and estimated \textit{fluence}. It also includes interactive links to flare light curves and AGILE AC-Lat4 data.


\section{Conclusions}

In more than 15 yr, from 2007 May 1st to 2022 August 31st, the AGILE AC system detected 5003 X-ray transients, compatible with a solar origin. A cross-check with data acquired by the GOES satellites in the same time interval allowed to confirm 3572 of them as genuine solar flares, and to retrieve their corresponding intensity class (i.e., B, C, M, and X). We also analyzed the RHESSI and Fermi GBM publicly available solar flares lists, to compare our results both with satellites operating in the SXR and HXR regimes.

The AGILE solar flare sample mostly consists in C class events, exhibiting a mean \textit{duration} of $13.1\pm0.3$~minutes, mean \textit{rise} times of $4.9\pm0.2$~minutes, and mean \textit{decay} times of $7.6\pm0.2$~minutes, and exhibiting average \textit{peak counts} of $(4.8\pm0.3)\times10^{3}$ and \textit{integrated counts} of $(2.7\pm0.2)\times10^{-6}$. These parameters are slightly different for each class of events, showing values in a general good agreement with those obtained in previous studies for the flares detected by GOES. A study of the offset between the flare maxima in the AGILE and GOES data highlights a longer persistence of the \textit{decay} phase in the AGILE HXR regime with respect to GOES. The same behavior is observed in the RHESSI and Fermi GBM solar flares, pointing out that such asymmetry between \textit{rise} and \textit{decay} phases is ascribed to the energy range. A study of the temporal offset between the occurrence of the flare \textit{maximum} in the GOES and in the AGILE data, shows a general tendency of the \textit{peak flux} to occur earlier in the SXR range, with respect to the HXR. Such behavior is also found in the RHESSI and Fermi GBM events, detected in the high X-ray regime. Both the longer persistence in the HXR and the temporal offset between the peaks observed in the SXR and in the HXR can represent further evidences that a two-phase mechanism can be involved in the acceleration of electrons in the solar atmosphere.

We studied the \textit{peak counts} and total \textit{integrated counts} distributions of the AGILE solar flare sample, for each associated GOES intensity class. Our events exhibit increasing \textit{peak counts} and \textit{integrated counts} with increasing intensity class. This behavior is completely compatible with what obtained in other solar flares lists, such as those of GOES, RHESSI, and Fermi GBM. Adopting a sample of GRBs detected by the AC-Lat4, whose \textit{fluence} is reported by other space missions, we provided a preliminary calibration factor to convert the AC counts to physical units. This passage allowed us to obtain an estimate of the \textit{peak fluxes} and the overall \textit{fluences} released by our events. Although AC data cannot be used to perform spectral analysis, this method allowed to carry out an approximation conversion from counts to physical units. The average flux of the AGILE solar flares obtained with this method in the 80-200~keV shows reasonable values, compared to the typical solar flare spectrum reported by RHESSI.

About one-fourth of the entire AGILE sample is constituted by transients not reported in the official GOES data and these events are therefore marked as unmatched solar flares. A cross-check with the new solar flares identified in the GOES data by using more effective detection algorithms, as well as with those reported by Fermi GBM and RHESSI, allows to confirm about 85\% of these transients as genuine events, mostly belonging to low-intensity, short-duration B- or C class events. The remaining 15\% can be ascribed to EOs suffered by geostationary satellites in 15 yr, and to the GOES satellites sensitivity and classification algorithm.

This study represents the first AGILE solar flare catalog, covering the end of the 23rd solar cycle, the whole 24th cycle, and the rising of the currently ongoing 25th cycle. This catalog provides a further database of solar flares in the HXR regime (80-200~keV), which can be used to integrate the other solar catalogs reported by other space missions, in order to carry out cross-checks and multiwavelength analyses. The results obtained in this study allow to better understand and characterize the capability of AGILE to carry out solar monitoring, and the sensitivity of the AC system to solar flares of different classes. Moreover, this work allowed to obtain a preliminary calibration of the AC system, which can provide preliminary reconstructions of the solar flares incoming \textit{peak flux} and \textit{fluence}.

AGILE continues its monitoring of the high-energy activity from the Sun with nominal capabilities, and we look forward to detect and collect a large number of further solar flares in the forthcoming research and promise more active phases of the current 25th solar cycle.


\acknowledgements

AGILE is a mission of the Italian Space Agency (ASI), with coparticipation of INAF (Istituto Nazionale di Astrofisica) and INFN (Istituto Nazionale di Fisica Nucleare). This work was carried out in the frame of the Addendum Nos.~6 and 7 - Accordo ASI-INAF n. I/028/12/0 for AGILE. This work makes use of data acquired by the Geostationary Operational Environmental Satellites (GOES), the Ramaty High-Energy Solar Spectroscopic Imager (RHESSI), and the Fermi Gamma-ray Burst Monitor (GBM), publicly available at \textit{https://data.nas.nasa.gov/helio/portals/} \textit{solarflares/datasources.html}.

\newpage

\bibliographystyle{aasjournal}       
\bibliography{bibliography_URSI.bib}   

\end{document}